\documentclass[12pt]{article}
\usepackage{epsfig,graphicx}

\def\be{\begin{equation}}
\def\ee{\end{equation}}
\def\bea{\begin{eqnarray}}
\def\eea{\end{eqnarray}}
\def\ba{\begin{array}}
\def\ea{\end{array}}
\def\re#1{(\ref{#1})}
\def\wt{\widetilde}

\newcounter{fig}

\begin{document}

\begin{titlepage}
\rightline{padthrev-mai05-f.tex}
\vskip 4 true cm
\centerline{
{\LARGE\bf{A Theory for the Term Structure}}
             }
\vskip 0.6 true cm
\centerline{
{\LARGE\bf{of Interest Rates}}
           }
\vskip 1 true cm
{\large\centerline{\bf Thomas Alderweireld{\small{\footnote[1]{
Thomas.Alderweireld@dexia.com, DEXIA group, square de Meeus, 1000 Brussels}}}
and Jean Nuyts{\small{\footnote[2]{Jean.Nuyts@umh.ac.be,
Universit\'e de Mons-Hainaut,
20 Place du Parc, 7000 Mons, Belgium}}}
}}

\vskip 2 true cm
{\noindent{\bf{Abstract}}}
The Convolution and Master equations governing the time behavior of the
term structure of Interest Rates are particularly simple for continuous
variables. The infinitesimal generator of the generalized Markov
process is usually a distribution.
We believe that
the discretised forms of the equation and of the generator are better suited to
compare with actual distributions. They help to avoid the Gaussian-like tail
behavior generally derived form the continuous equations with a finite number of
Markov processes. In this paper, the notion of discretised Seed is
introduced
which naturally leads to an infinite superposition of Markov processes and hence
allows a power (rather than exponential) decrease of the empirical probabilities
of the variations of the interest rates. The discretised
theoretical distributions of probabilities
matching the empirical data from the Federal Reserve System (FRS) are deduced
from a discretised seed which enjoys remarkable scaling laws. In particular the
tails
of the distributions are very well reproduced. These results may be used to
develop new methods
for the computation of the value-at-risk and fixed-income derivative pricing and
suggest the appearance of the critical exponents related to models based on
self-organized systems.

\end{titlepage}
\eject
\setcounter{page}{1}

\section{Introduction \label{sec:Intro}}

The accuracy of the  interest rates variations modelling is an
important issue especially in the context of  the evaluations of the value-at-
risk and marked-to-market positions in trading floors.

In two recent articles \cite{NP}, \cite{AN}, it was shown using empirical data
published by the governors of the Federal Reserve System \cite{FRS}
from 1962 until 2002 that the term structure of interest rates decreases
essentially for large variations of the  interest rates
as a power  and that this power is of the
order three
to four. Moreover the distributions seem to
obey simple approximate scaling laws as functions of the initial interest
rate, of the lag and of the maturity.
These findings invalidate many models which predict distributions having either
very short tails, generally
exponentially decreasing, or very long tails as do Levy type structures
\cite{L}, \cite{M1}, \cite{M2}, \cite{WWW}.

In this paper, a theoretical model is built to
serve as a basis for computing the distribution of the variation of interest
rates in terms of a few fundamental ``microscopic parameters'' whose
meaning will be highlighted. At the basis of the theory, the notion of "seed" (a
discretised form of the infinitesimal generator \cite{BRP})
is introduced. It is closely related to the variation
of the interest rates for a very short but finite time intervals. We believe
that this discretisation is an important issue, closely linked to the fact that
empirical interest rates are, often by regulation, expressed as integer
multiples of the basis point and obtained by averages performed on a finite set
of discrete times.

In order to simplify the presentation
and highlight preceding results available in the litterature \cite{BRP}, the
problem is set
in terms
of continuous variables. Later, to allow numerical simulations and come
closer to the empirical distributions, the variables are discretised.

All along this article, in order to make a connection to a real
situation,  we have
chosen to
refer systematically to an application of our ideas to the FRS data \cite{FRS}.
Needless to say, we expect our analysis to be extendable, mutatis mutandis, to
many
other situations.

\section{The basic equations with continuous variables \label{sec:contbasic}}

In this section, the basic equations, which govern the
continuous time
propagation of the term structure of interest rates, are briefly reviewed,
analysed and commented from a econo-physicist's point of view.

Suppose that, the interest rate for a certain maturity $[m]$ has the value $I_0$
at time $t_0$. We want to study the normalized density of probability
(as seen at time $t$)
\be
p_{t}^{[m]}(t_f,I_f,t_0,I_0)
\label{p1}
\ee
that,
at a later (final) time $t_f$, thus after a lag
\be
L=t_f-t_0\ ,
\label{lagdef}
\ee
the interest rate
has a value $I_f$. In principle, this density of probability $p_t$ is an
unknown
function of the five ``continuous variables'' $t,t_f,I_f,t_0,I_0$. In order to
simplify the notation we will restrict ourselves to a given maturity and
suppress the corresponding upper index.

The compounded probability (at time $t$)
$\overline{P}_{t,t_f,I_a\leq I_f\leq I_b,\,t_0,I_0}$ that, starting with an
initial interest rate $I_0$ at the
initial time $t_0$, the final interest rate $I_f$ at the final time $t_f$ is in
the interval between $I_a$ and $I_b$ is given by the integral
on this interval of the
probability density
\be
\overline{P}_{t,t_f,I_a\leq I_f\leq I_b,\,t_0,I_0}=
      \int_{I_a}^{I_b} p_t(t_f,I_f,t_0,I_0)\, dI_f
\label{prob}
\ee
with, obviously, by normalization
\be
\overline{P}_{t,t_f,-\infty\leq I_f\leq +\infty,\,t_0,I_0}=1\ .
\label{prob2}
\ee

We will now make the hypothesis that the
interest rate variations satisfy
some market laws which are rather stable and are governed by sufficiently
smooth
equations. Let us try to state more precisely and justify our simplifying
assumptions.

\subsection{Time translation invariance\label{s:timetr}}

This is a very delicate hypothesis. It is equivalent in
saying that, whatever be the political or economical situation the average
behavior will be identical. The effects of the exceptional situations
(political or economical crisis) which usually may lead to seemingly
incoherent
variations of the interest rates are accurately taken into account by the tails
of the functions which are used. The postulate is that, even if there
are extreme situations of various importance, all in all they connect
smoothly
with the more normal situations which prevail during the peaceful times. It
is
precisely these extreme situations which are the prime reason for the fat
tails
of the distributions. And they must be incorporated correctly by the
model. To repeat,
the hypothesis is that what happens in the exceptional situations is
well taken into account by
the smooth tails of the distributions. There is no abrupt transition
between really exceptional situations and what we would call the normal
situations. Between the extremes, there are situations of intermediate
seriousness
which lead for the distribution to a smooth passage
from a restful period to a chaotic one. But let us make our arguments more
precise.

Mathematically,
for any of the times $t,t_0,t_f$, the global time translation operation is given
by
\bea
t'&=&t+T
\label{timetr}\\
t'_0&=&t_0+T
\label{timetr0}\\
t'_f&=&t_f+T   \ .
\label{timetrf}
\eea
where $T$ is the value of the translation time (minutes or days or months
or
years later).
Suppose that at some time $t'_0$ later than $t_0$ given by the translation
time $T$
the interest rate $I'_0$ is again exactly equal to $I_0$.
Technically, the time translation invariance demands that
the densities of
probability to obtain the same final interest rate $I'_f=I_f$ be equal
\be
p_{t'}(t'_f,I_f,t'_0,I_0)=p_t(t_f,I_f,t_0,I_0)
\label{timetrP1}
\ee
whatever be $T$.
The financial interpretation of this equality is that the market laws are stable
in time. If the same situation is reproduced later, it will evolve with the same
probabilities.

It is not difficult to prove mathematically that this condition essentially
implies that the density of probability does not depend on the three time
variables $t,t_0,t_f$ separately but only on two of their differences, say
$t-t_0$ and the lag
\be
L=t_f-t_0 \ .
\label{lag}
\ee
Let us finally discuss the dependence in $t$ which is the time at which the
information is needed. Obviously if $t$ is larger than $t_0$, in principle all
the information of the what the actual interest rates have been between $t_0$
and $t$ is available and known.	The related probabilities are nor useful any
more. If $t$ is smaller than $t_0$ it means that we want to evaluate the
probabilities for evolutions of the interest rates happening at later times.
Again, if the market laws are stable, the evolution during a fixed lag $L$ as
seen at the time($t=t_0$) or at a earlier time ($t<t_0$) should be identical.
The same initial state ($I_0$ at $t_0$) should lead
statistically to the same final state ($I_f$ at $t_f$). This means invariance
under a further logically different time transformation \re{timetr} while $t_0$
and $t_f$ are kept fixed
\bea
t'&=&t+T
\label{timetrspe}\\
t'_0&=&t_0
\label{timetr0spe}\\
t'_f&=&t_f
\label{timetrfspe}
\eea
This implies that the density of probability
is independent of its index $t$.

Gathering the result of the two invariances, we find the time invariant
restricted form of the density
\be
p_t(t_f,I_f,t_0,I_0) = {\widetilde{p}}(t_f-t_0,I_f,I_0)  \ .
\label{timetrP2}
\ee
In other words, it does not depend independently
on $t$, $t_f$ and $t_0$ but on the lag only.

The invariance under the two continuous time translations thus implies that
the density of probability $p_t$ is reduced to a function
of three continuous variables only: the lag,
the final interest rates $I_f$ and initial interest rate $I_0$.
The variation
$V$ of the interest rate during the lag $L$ is defined as
\be
V=I_f-I_0\ .
\label{var}
\ee
For later convenience, a last variable change is performed to define the
time translation invariant density of probability $p$ as a function of the lag,
of the initial interest rate and of the variation of the interest rate.  From
now on, the form
\be
p(L,V,I_0)\equiv {\widetilde{p}}(t_f-t_0,V+I_0,I_0)
\label{basicdensy}
\ee
will be used as the basic interest rate distributions.

\subsection{The normalization of the probability \label{s:norma}}

The total probability \re{prob2} that after the lag $L$ the rate $I_f$ has any
value
must be equal to one. This implies the normalization
\be
\int_{-\infty}^{+\infty}p(L,I_f-I_0,I_0)\, dI_f\equiv
\int_{-\infty}^{+\infty}p(L,V,I_0) \, dV=1
\label{norm}  \ .
\ee
This normalization should hold whatever be the lag and whatever be the
initial
rate.

\subsection{The composition of the probabilities\label{s:compose}}

The basic equations, which govern the composition of
the
probability densities $p$
(essentially the Chapman-Kolmogoroff equations \cite{BRP}), are well-known.
Let us write them in our notation.
Intuitively, consider three times
$t_0,t_i,t_f$
where
the
intermediate time $t_i$ lies between the initial time $t_0$ and the final time
$t_f$.
\be
  t_0\leq t_i \leq t_f    \ .
\label{intertime}
\ee
The initial rate is $I_0$.

During the time lag $L_1=(t_i-t_0)$ the interest rate has a density of
probability
$p(t_i-t_0,I_i-I_0,I_0)$ to reach the intermediate value $I_i$. Then starting
from the intermediate time
$t_i$ up to final time $t_f$ (i.e. during the second time lag $L_2=(t_f-t_i)$)
the
intermediate observed interest
rate $I_i$ at time $t_i$ has a density of probability $p(t_f-t_i,I_f-I_i,I_i)$
to become
$I_f$
at the
final time $t_f$. Since the interest rate at the intermediate time $t_i$ can
take any
value,
the density of probability starting from the rate $I_0$ at initial time $t_0$ to
end up
with a rate
$I_f$ at the final time $t_f$ is given by the integration on $I_i$ at the
intermediate time
(convolution of the probabilities)
\be
p(t_f-t_0,I_f-I_0,I_0)=
\int_{-\infty}^{+\infty}p(t_f-t_i,I_f-I_i,I_i)\,
                        p(t_i-t_0,I_i-I_0,I_0)\, dI_i  \ .
\label{convolution}
\ee
This is the basic equation which the probability distribution has to fulfil.
It shows how the probability distributions of two successive lags
$L_1=(t_i-t_0)$ and $L_2=(t_f-t_i)$
compose to form the probability distribution for the lag $L=t_f-t_0=L_1+L_2$.
Eq.\re{convolution} should hold whatever be the intermediate time.

It is convenient to rewrite the equation by using the new variables $I,V$
and
the new integration variable $W$
\be
\ba{rclcrcl}
I&=&I_0&\quad,\quad& I_0&=&I
   \nonumber\\
V&=&I_f-I_0&\quad,\quad&I_f&=&V+I
\nonumber \\
W&=&I_i-I_0&\quad,\quad&I_i&=&W+I
\nonumber \\
L_1&=&t_i -t_0&\quad,\quad&L_2&=&t_f-t_i
\ea
\label{newvar}
\ee
as
\be
p(L_1+L_2,V,I)=
\int_{-\infty}^{+\infty} p(L_2,V-W,I+W)\, p(L_1,W,I)\, dW
\label{convolnew}
\ee
This is the basic equation.

\subsection{Initial conditions\label{s:init}}

The probability distribution has to satisfy an initial condition which can be
described in terms of the Dirac $\delta$ distribution (see Appendix A).
Indeed
if the initial
rate at time $t_0$ has the value $I_0$ (whatever $I_0$ is), at $t_f=t_0$ (i.e.
after a zero lag
$L=t_f-t_0=0$) we know for sure that the rate is still $I_0$. The density of
probability for the final rate to be different from $I_0$ is zero. In other
words, when the lag is
zero, the rate has not moved. It is still $I_0$ with probability one.

Mathematically, this implies that
\be
p(0,V,I_0)=\delta(V) \ .
\label{deltaprob}
\ee
Some properties of the distribution $\delta(V)$ are given in Appendix
\re{ap:delta}
together with a few useful approximations in terms of more conventional
functions which will be used later. In particular, the Dirac distribution is
normalized
\be
\int_{-\infty}^{+\infty}p(0,V,I_0)\, dV
\equiv
\int_{-\infty}^{+\infty}\delta(V)\, dV=1
\label{deltanormp}  \ .
\ee
As it should, in agreement with the composition of probabilities
\re{convolution} evaluated either for $t_i=t_0$ or for $t_i=t_f$, the Dirac
distribution, composed with
any distribution, satisfies the identities
\bea
p(t_f-t_0,I_f-I_0,I_0)&=&
\int_{-\infty}^{+\infty}p(t_f-t_0,I_f-I_i,I_i)\,
                        p(0,I_i-I_0,I_0)\, dI_i
             \nonumber\\
                        &=&
\int_{-\infty}^{+\infty}p(0,I_f-I_i,I_i)\,
                        p(t_f-t_0,I_i-I_0,I_0)\, dI_i
\label{deltaconv}
\eea
showing again that an evolution during a zero lag is, in fact, not an
evolution.

\subsection{The seed \label{s:seed}}

By using the convolution equation \re{convolution}, if one knows the probability
distributions $p(\epsilon,V,I_0)$ for a given lag $L=\epsilon$, whatever be the
value of $\epsilon$, the
probability distribution for $p(2\epsilon,V,I_0)$ for a lag of $2\epsilon$
can easily be computed
by the convolution of $p(\epsilon,V,I_0)$ with
itself. Then $p(3\epsilon,V,I_0)$ is obtained by the convolution of
$p(2\epsilon)$ with $p(\epsilon,V,I_0)$.
By successive iterations $p(n\epsilon,V,I_0)$ can be computed for any positive
integer $n$. Hence, if one knew the distribution
of probability
for a very small lag, $\epsilon$ much smaller than the empirical lags,
the distribution
for a
lag $L$ much larger than $\epsilon$
could be computed approximately
by simple successive integration. Essentially a number of
integration equal to the integer $n$ closest to $(L/\epsilon-1)$. This
approximation
would become better and better when $\epsilon\rightarrow 0$.

It is then tempting to let $\epsilon$ go to zero and to take the first order
approximation, which we call the seed $S(V,I_0)$ for reasons to follow, as
\bea
S(V,I_0)&=&\lim_{\epsilon\rightarrow 0}
       \frac{p(\epsilon,V,I_0)-p(0,V,I_0)}{\epsilon}
     \nonumber \\
     &=& \partial{_{\lower0.5ex\hbox{\scriptsize{\it{L}}}}}\, p(L,V,I_0)\,
        {\vrule height11pt width0.4pt depth 10pt}_{L=0} \ .
\label{seed}
\eea
Since the $V$-integration of the two $p$ terms in the right hand side are
equal
(before $\epsilon$ is put to zero) by \re{norm}, the seed $S$ satisfies the
normalization condition
\be
\int_{-\infty}^{\infty} S(V,I_0)\, dV=0   \ .
\label{normS}
\ee

This seed is essentially related to what is often called the "infinitesimal
generator" of a Markov process in analogy with the infinitesimal generator
familiar in the theory of Lie groups and semi-groups. It allows, in principal,
the computation of the time evolution for a finite time from the knowledge of
what has happened during an infinitesimal small amount of time.

We have however decided to call it seed for two reasons. First, all the
variables (interest rates and times) are discretised. Thus, the meaning of the
seed does not coincide with the meaning of the infinitesimal generator as it
does not correspond to an infinitesimal time anymore. Second, Markov processes
are usually based on random walk models. For a finite superposition of Markov
processes (arbitrary large but finite) the distribution of probability turns out
to be a finite superposition of exponentials (short tails)
(see \re{sec:master2}). Our generalisation
will in fact be equivalent to  an infinite superposition and hence will
transcend exponentials and hence escape the ill-fated short tails.

If a Taylor expansion is used, in first approximation, for $L=\epsilon$ very
small,
the probability distribution $p$ can written
\be
p(\epsilon,V,I_0)=p(0,V,I_0)+\epsilon S(V,I_0)  \ .
\label{taylor}
\ee
In the continuous case, this is a distribution. Nevertheless in the discretised
case, this form, with function rather distributions, can be used as a very good
approximation.

In the rest of the section, we discuss a few facts related to the presence of
distributions rather than functions. The reader which is not interested in these
mathematical considerations should jump immediately to section \re{sec:num}.
Indeed, technically, it should be stressed that, as the initial condition
\re{deltaprob}
is a distribution rather than a function,
we should expect that the seed is also a
distribution with support restricted to $V=0$.

\subsection{The Master Equation. Continuous variables \label{s:master}}

If the seed is known as a distribution, the density of probability of the
interest rate variation
can easily
be computed by integrating the master equation as an integro-differential
equation
\be
\partial_{\lower0.5ex\hbox{L}}\  p(L,I_f-I_0,I_0)=
\int_{-\infty}^{+\infty} p(L,I_f-I_i,I_i)\, S(I_i-I_0,I_0)\, dI_i
\label{master}
\ee
which follows from the convolution equation \re{convolution}.
With the change of variables \re{newvar}
the master equation can equivalently be written
\be
\partial_{\lower0.5ex\hbox{L}}\  p(L,V,I)=
\int_{-\infty}^{+\infty} p(L,V-W,I+W)\, S(W,I)\, dW \ .
\label{masternew}
\ee

Technically, one obtains Eq.\re{master} by differentiating the convolution
equation
\re{convolution} with respect to $t_i$ and letting $t_i\rightarrow t_0$.
The master equation \re{master} is a consequence of the convolution
equation. Conversely, in Appendix \re{ap:formal},
it is shown, formally, that the solutions of the master
equation satisfy the convolution equation \re{convolution}.

\section{The Gauss distribution as the solution for the simplest seed
\label{sec:master2}}

The simplest possible \re{deltasecond} (see Appendix \re{ap:delta}) seed is the
distribution
\be
S(V,I_0)=\kappa(I_0)\,\partial_V^2\,\delta(V)  \ .
\label{seeddelta}
\ee
This can be justified na\"\i vely as follows. During a very small amount of time
$\epsilon$ one expects the variation of the interest rate $V$ to be very small,
say at most of the order $\rho$ where $\rho$ decreases and goes to zero with
$\epsilon$.
Introduce the three intervals of length $\rho$, the left interval $L=
[-3\rho/2,-\rho/2]$, the center one $C=[-\rho/2,+\rho/2]$ and the right one
$R=[+\rho/2,+3\rho/2]$. Outside these three intervals the probability is
essentially zero as is the density of probability. In the left and right
intervals the probability is small say of the order $\kappa$ and in the center
it must be $1-2\kappa$ to conserve the normalization of the probability. If the
density of probability is then supposed to be constant in the three intervals,
the
seed becomes
$\kappa/\rho$ in the left and right intervals and $-2\kappa/\rho$ in the center
interval. Letting the time interval $\epsilon$ go to zero induces $\rho$ to go
to zero and the limiting seed becomes, up to $\kappa$, the second derivative of
the $\delta$ distribution. A Markov process is a {\it{finite}} superposition of
terms
of the form \re{seeddelta} with possible higher (even) derivatives of
the $\delta$ distribution.

Taking this simplest form of the seed, the integro-differential equation
\re{master} becomes a much simpler
partial differential equation
\be
\partial_{\lower0.5ex\hbox{\scriptsize{\it{L}}}}\,p(L,I_f-I_0,I_0)
      =\kappa(I_0)\,\partial^2_{I_0}\,p(L,I_f-I_0,I_0)
           \nonumber\\
\label{master2}
\ee
which is also written
\be
\partial_{\lower0.5ex\hbox{\scriptsize{\it{L}}}}\,p(L,V,I_0)
      =\kappa(I_0)\,
      \biggl(\partial^2_{V}\,p(L,V,I_0)
             -2\,\partial_V\,\partial_{I_0}\,p(L,V,I_0)
             +\partial^2_{I_0}\,p(L,V,I_0)\biggr)   \ .
\label{master2b}
\ee
This equation is still rather complicated.

It is shown in Appendix \re{ap:gauss}, assuming no $I_0$ dependence and
simplified scaling laws \re{mas4dista} and \re{mas4distb}, that the solution of
the master equation for the seed \re{seeddelta} is of Gaussian type
\be
\wt{p}({\wt{V}})
=\frac{1}{\sqrt{2\pi\sigma^2}}\,e^{-\frac{\wt{V}^2}{2\sigma^2}}  \ .
\label{master4solcc}
\ee

This solution as highlighted in \cite{NP} and \cite{AN} is completely excluded
by the observations as are all Markov's approaches which lead to finite
superpositions of such Gaussians.

\section{Some considerations about the discretisation of the problem.
Numerical integration\label{sec:num}}

\subsection{Why discretisation? {\label{whydisc}}}

The choice of discretising the distributions is not done here for convennience
but is motivated by the discreet nature of the available data and the way they
are produced by the laws of supply and demand. Naturally, market makers are
always working in units of ``something´´, $1/32$ of \$, basis point $\ldots$.
Thus, the natural variation measure units are by nature discreet. The same
discussion applies to
Federal Reserve System data \cite{FRS}. In this case, we are quite
naturally led to discretise both the time variable and the interest rates
variable. As a consequence the master equation be better discretised
Indeed, first, by regulation, the interest rates are expressed by integers in
basis point ($10^{-2}$ percent) and hence naturally discreet. Then, usually, the
determination of the daily average interest rates is compounded by a finite set
of individual contracts finalized at certain times. These times do not occur at
fixed intervals. But they are clearly not continuous and probably better taken
into account by a set of discrete times at regular intervals rather than by a
continuous time. This is a critical point. We show that the discretisation leads
to extremely accurate results.

Moreover, technically, given a seed which is a distribution, the convolution
equation \re{convolnew} or the master
equation \re{masternew} are usually very difficult or impossible to solve
analytically. Instead, in this present work, a
numerical procedure has been implemented.
This numerical integration depends on the discretisation of the
continuous equation using discretised variables on a grid.

We have focussed our attention to
the following specific form of the
continuous equations \re{convolnew} for $L=L_1=L_2$
\be
p(2L,V,I)=
\int_{-\infty}^{+\infty} p(L,V-W,I+W)\, p(L,W,I)\, dW  \ .
\label{convolnewss}
\ee

In the following, the ideas and the delicate points are briefly described.
The precise application of the formalism to the FRS data \cite{FRS} and
the results are summarized and highlighted in section
\re{sec:results}.

In this section are presented in turn the ideas which are needed to perform the
approximate numerical integration of the equation. The continuous equations are
discretised by associating a finite lattice (a grid) to the continuous
variables. The initial conditions and the seed are then defined on a finite set
of
points at the nodes of the grid. This allows then to define the $\chi$-squared
merit function.

\subsection{The ideas behind the grid \label{s:grid}}

The equation \re{convolnewss} depends on three continuous variables. In order to
perform numerical integrations, the three variables have to be discretised.

\begin{description}

\item{The Lag grid}

Suppose that the distribution is known for some value $s_L$ of the lag $L$. From
the first application of \re{convolnewss} one obtains the distribution for
$L=2s_L$. Applying \re{convolnewss} a second time allows the determination of
the
distribution for $4s_L\equiv 2^2 s_L$. The iteration of the procedure produces
the
distribution for any $L$ of the form $2^ls_L$ i.e. on a the $L$-grid of values
\be
L=2^l\cdot s_L \quad l=0,\ldots,N_L   \ .
\label{Lgrid}
\ee

\item{Our choice of the Lag grid}

We have chosen to obtain a lag $L=1$ day after $N_L=10$ iteration. This leads to
a
value of
\be
s_L=\frac{1}{2^{10}}\ {\rm{days}}
\label{sL}
\ee
i.e. about 30 seconds. This is a perfect choice
if
individual contracts are concluded at this rate. If it turns out that the lapse
between two successive contracts is smaller, $N_L$ should be increased.

\item{The $V$ grid}

For the $V$ variable, the best choice is a regularly spaced grid consisting of
points of the form
\be
V=v\cdot s_V
\label{Vgrid}
\ee
where $s_V$ is the step size in $V$ and $v$ is an integer. A smaller $s_L$
should lead to a choice of a smaller $s_V$.

\item{Our choice of the $V$ grid}

We have chosen to take $s_V$ to be
\be
s_V=\left(\frac{1}{1.4}\right)^{10}\ {\rm{basis\ point}}
\label{sV}
\ee
i.e. around
$(\frac{1}{\sqrt{2}})^{10}$ basis point. Indeed a larger $s_V$ turned out not to
be small enough while a
smaller one was not necessary and increased the computation time and memory
requirements with little
reward.

\item{The $I_0$ grid}

For the $I_0$ variable, the best choice is a regularly spaced grid consisting of
points of the form
\be
I_0=i\cdot s_I
\label{Igrid}
\ee
where $s_I$ is the step size in $I_0$ and $i$ is an integer.

\item{Our choice of the $I_0$ grid}

Since all the parameters vary with $I_0$ rather weakly, a choice of
\be
s_I={\rm{one\ percent}}
\label{sI}
\ee
is quite sufficient.

\end{description}

\noindent In our actual computations due memory size problems, we have been led
to adapt the $s_V$ step size to the level $l$ of the iteration. The step size is
progressively increased in such a way as to become one basis point at the tenth
iteration. This required delicate numerical adjustments

\subsection{The ideas behind the discretised equations \label{s:discequa}}

Obviously, on the grid \re{Lgrid}, \re{Vgrid}, \re{Igrid} the equation
\re{convolnewss} becomes
\bea
p(2^{l+1} s_L,v\,s_V,i\, s_I)&\approx &s_V
	\left\{ \sum_{w=-\infty}^{\infty}
	\right.
	p\left(2^{l} s_L,(v-w) s_V,i\, s_I+w\, s_V\right)
	\times
	\nonumber\\
	&&
	\phantom{{V}\sum_{w=-\infty}^{\infty}}
	\left.
	\times\
	p\left(2^{l} s_L,w\, s_V,i\, s_I\right)
	\right\}    \ .
\label{disconvolss}
\eea
To be on the grid, the argument $i\, s_I+w\, s_V$ which appears
in the first $p$ on
the right hand side must be an integer multiple of $s_I$. This would at
first sight imply that
$s_V$ is an integer multiple of $s_I$.
In fact, our numerical implementation
uses, for technical reasons, a more elaborate grid than
\re{Vgrid} in the $V$ and a better adapted form of the convolution
\re{disconvolss} together with smoothing and interpolation techniques.
Moreover the
normalization implies
\be
\sum_{v=-\infty}^{\infty}
	p\left(2^l s_L,v\, s_V,i\, s_I\right)=1   \ .
\label{discnorm}
\ee

\subsection{The ideas behind the discretised initial condition
\label{s:discini}}

The initial condition \re{deltaprob} on the grid
is taken as a step-type \re{stepdelta} approximation of the Dirac $\delta$
distribution, as explained in Appendix \re{ap:delta}.
Namely \bea
p(0,0,i\,s_I)&=&\frac{1}{s_V}\quad ,
    \nonumber\\
p(0,v\,s_V,i\,s_I)&=&0\quad\quad{\rm{for}}\ v\neq 0   \ .
\label{inigrid}
\eea

\subsection{The ideas behind the discretised seed \label{s:discseed}}

Using the discretisation initial condition, for $\epsilon=s_L$ chosen
sufficiently small, the Taylor
expansion \re{taylor} of $p$ to
the first order in $s_L$
\be
p(s_L,v\,s_V,i\,s_I)=p(0,v\,s_V,i\,s_I)+s_LS(v\,s_V,i\,s_I)   \ .
\label{distaylor}
\ee
The resulting $p(s_L,v\,s_V,i\,s_I)$ must be a true function (not a
distribution anymore). It must be positive and normalized \re{discnorm}
\be
\sum_{v=-\infty}^{\infty}
	S\left(v\, s_V,i\, s_I\right)=0\ .
\label{discnormS}
\ee
Hence, $S(v\,s_V,i\,s_I)$ is expected to be a step-type function of $vs_V$
with the
following properties
\bea
S\left(0,i\, s_I\right)&=&- \sum_{v\neq 0}S\left(v\, s_V,i\, s_I\right)
    \nonumber\\
S\left(v\, s_V,i\, s_I\right)&>&0 \quad\quad{\rm{for}}\ v\neq 0
\label{disposS}
     \\
S\left(0,i\, s_I\right)&<&\frac{1}{s_Ls_V}   \ .
    \nonumber
\eea
The form of the seed inferred by the FRS data is presented and discussed in
section~\re{sec:FRS}.

\subsection{The ideas behind the $\chi$-squared function \label{s:chisquared}}

In order to determine the seed free parameters, a $\chi$-squared
minimization method is used.

As usual the generic $\chi$-squared function is defined by
\be
\chi^2=\sum_k\frac
          {\left(N_{\rm{theory}}(k)-N_{\rm{data}}(k)\right)^2}
		{\sigma(k)^2}
\label{chicarre}
\ee
where the sum is performed on the generic discrete label $k$ indexing all the
available data. In \re{chicarre},
$N_{\rm{data}}(k)$ and $N_{\rm{theory}}(k)$ are respectively the number of
events observed and predicted for the label $k$.

The natural error ($\sigma(k)_{\rm{natural}}$) is usually taken as if the
distribution was of Poisson type i.e.
\bea
\sigma(k)_{\rm{natural}}&=&\sqrt{N_{\rm{data}}(k)}\quad\quad {\rm{if}}
			\ N_{\rm{data}}(k)\neq 0
        \nonumber\\
\sigma(k)_{\rm{natural}}&=& 1 \quad\quad\quad \quad\quad
			{\ \ \rm{if}}\ N_{\rm{data}}(k)= 0   \ .
\label{sigmai}
\eea

The choice of the error function requires more attention in the present case.
As has been already stressed,
the theory should take into account and reproduce as accurately as possible the
tails of the experimental distributions. This is achieved by choosing in the
$\chi$-squared a modified form of $\sigma(k)$
\bea
\sigma(k)_{\rm{modified}}
   &=&\left(N_{\rm{data}}(k)N_{\rm{theory}}(k)\right)^{1/4}
   			\quad\quad\ {\rm{if}}
			\ N_{\rm{data}}(k)\neq 0
\nonumber\\
\sigma(k)_{\rm{modified}}
   &=&\left(N_{\rm{theory}}(k)\right)^{1/4}
   			\quad\quad\quad\quad\quad\quad {\rm{if}}
			\ N_{\rm{data}}(k)= 0   \ .
\label{sigmamod}
\eea
Obviously, if theory and experiment match very closely the $\sigma$'s in
\re{sigmai} and \re{sigmamod} are essentially identical. This happens in the
bulk of the distribution. For the tails of the
distributions, when $N_{\rm{data}}$ is equal to zero or one event, the
$\sigma(k)_{\rm{modified}}$ is smaller than $\sigma(k)_{\rm{natural}}$. Hence,
more emphasis is put on these tail points by the minimization procedure. The
tests
performed using both forms of $\sigma(k)$ confirm this argument. The use of
\re{sigmamod} leads to a very good agreement between theory and data in the
tails as can be seen in the Figures.

\section{The Federal Reserve System data. Determination of
$N_{\rm{data}}(v,i_{\rm{bin}})$ \label{sec:FRS} for each allowed maturity}

The FRS \cite{FRS} data gives in successive working days the daily average
interest rate $I({\rm{day}})$ between banks. They are given in
Table \re{eventsmat}for the eight
following maturities
\be
\{[m]\}=\{[1],[2],[3],[5],[7],[10],[20],[30]\}\ \ {\rm{in\ years}}\ .
\label{matset}
\ee
For many of
these maturities, the data extends (10698 events) without break from 1962 to
today (we have
chosen November 4 as our last event). For a few maturities, the data is
restricted to one or more sub-periods within the 1962-today period. The total
number of events for each maturity is given in table \re{eventsmat}.
In particular, it should be stressed that the data for $[m]=20$ years consist of
two disjoint periods. We have not been able to discover if, during theses two
periods, the way of computing the daily averages are coherent. We have chosen to
group the two sets but we will be led to caution comments.

\begin{table}
\vskip 0.5 true cm
\begin{center}
\begin{tabular}
{|c|c|c|c|c|c|c|c|c|c|c|c|c|c|c|c|c|c|}
\hline
$[m]$ &$[1]$ &$[2]$ &$[3]$ &$[5]$ &$[7]$ &$[10]$ &$[20]$ &$[30]$   \\
\hline
$N_{eve}$ &10698 &7106 &10698 &10698 &8828 &10698 &9009 &6243 \\
\hline
\end{tabular}
\end{center}
\caption{The number of events $N_{eve}$ for the maturities $[m]$.
\label{eventsmat}}
\end{table}

The interest rates are given in percent by a number with exactly
two decimal
figures, i.e. by a integer number in basis points. It allows the definition of
the
following meaningful distributions for a lag $L$ of one day. These distributions
have
to be compared to the corresponding theoretical distributions.

\begin{description}

\item{\bf{The empirical $N(v,I)$ and $\overline N(v,i)$ density distributions}}

For $v$ and $I$ both expressed in basis points and for each of the allowed
maturities, the empirical distribution $N(v,I)$
is the number of occurrences when the interest rate of
the FRS on some day was $I$ and the next day $I+v$. These numbers are
statistically very
small. It is useful to consider the compounded empirical distribution
$\overline N(v,i)$ where $v$ is still expressed in basis points but the interest
$i$ is expressed in percent by grouping the days when the interest rate is
almost $i$

More precisely, for the data, the $N(v,I)$ ($v$ and $I$ in basis points) density
distribution is defined as follows
\be
N(v,I)={\rm{number\ of\ days\ when\ }}I({\rm{day}})=I
	{\rm{\ and\ }}I({\rm{day}}+1)=I+v\ .
\label{density}
\ee
Note that, in order to obtain $I({\rm{day}}+1)$, non working days are simply
discarded.
The average discretised
$\overline N(v,i)$ ($v$ in basis points and $i$ in percent)
density distribution is precisely defined by
\be
\overline N(v,i) =\sum_{I=100i-49}^{100i+50}N(v,I)\ .
\label{densitydisc}
\ee
It is the number of days when the interest rate was between $i-1/2$ percent (
exactly
$100i-49$ basis points) and
$i+1/2$ percent (exactly $100i+50$ basis points)
and the interest rate has moved by an amount $v$ basis points by the next day.

\item{\bf{The empirical $w(I)$ and $\overline{w}(i)$ interest
distributions}}

For each allowed maturity,
the empirical interest distribution $w(I)$ is defined as
\be
w(I)=\sum_{v_{min}}^{v_{max}}N(v,I)
\label{Idist}
\ee
where $I$ is in basis point. It is the number of days when the interest rate was
$I$ in
basis point. The average discrete version $\overline{w}(i)$ of \re{Idist} is
defined by
\be
\overline{w}(i)=\sum_{I=100i-49}^{100i+50}w(I)   \ .
\label{Idistdisc}
\ee
where $i$ is an integer giving the interest rate in percent. it is simply the
number of
days when the interest rate was between $(i-1/2)$ and $(i+1/2)$ percents.
The total number of events $\overline{w}=\sum_i\overline{w}(i)$ depends on the
maturity and is given in Table \re{eventsmat}.

The $\overline{w}(i)$ are given in Table \re{t:w1} for all
maturities of $[m]$ years. In preceding papers
it was realized that bins with less than about one thousand events present too
much
statistical fluctuations. Hence it is useful to group the data in bins.
We have chosen this bin definition identical for
each of the allowed maturities.
In Table \re{t:wbin}, the $\overline{w}(i_{\rm{bin}})$ are given, as an example,
for
maturity $[m]=1$, together
with a definition of the bins.

\begin{table} 
\vskip 0.5 true cm
\begin{center}
\begin{tabular}
{|c|c|c|c|c|c|c|c|c|c|c|c|c|c|c|c|c|c|c|}
\hline
$i$ &1&2&3&4&5&6&7&8&9   \\
\hline
$\overline{w}(i)_{[1]}$ &356 &402&664&1269&1674&2147&1110&1147&582 \\
\hline
$\overline{w}(i)_{[2]}$& 60&418&293&569&599& 1530&759&835&561 \\
\hline
$\overline{w}(i)_{[3]}$& 9&369&524&1246&1254&2166&1642&1334&732 \\
\hline
$\overline{w}(i)_{[5]}$ &0&33&45&130&1278&2010&1791&1563&778 \\
\hline
$\overline{w}(i)_{[7]}$&0&0&154&495&693&1705&1775&1582& 880 \\
\hline
$\overline{w}(i)_{[10]}$& 0& 0&28&1485&1246&1949&1874&1629& 929 \\
\hline
$\overline{w}(i)_{[20]}$&0&0&0&1020&1320&1974&1389&1341& 396 \\
\hline
$\overline{w}(i)_{[30]}$& 0& 0& 0& 0&289&1076& 968&1358&981 \\
\hline
\end{tabular}
\end{center}
\caption{The empirical $\overline{w}(i)_{[m]}$ for $i=1-9$
and all maturities $[m]$.\label{t:w1}}
\end{table}

\begin{table} 
\vskip 0.5 true cm
\begin{center}
\begin{tabular}
{|c|c|c|c|c|c|c|c|c|c|c|c|c|c|c|c|c|c|c|}
\hline
$i$ &10&11&12&13&14&15&16&17&18 \\
\hline
$\overline{w}(i)_{[1]}$&469&176&202&105&160&131&65&38&0    \\
\hline
$\overline{w}(i)_{[2]}$& 445& 294& 196& 157& 190&126&55&18& 0 \\
\hline
$\overline{w}(i)_{[3]}$&333&351&179&173&234& 88& 59&  4&  0 \\
\hline
$\overline{w}(i)_{[5]}$& 341& 338& 222& 216& 240&94&39&0&0 \\
\hline
$\overline{w}(i)_{[7]}$& 332& 306& 300& 250& 233&96&26&0&0 \\
\hline
$\overline{w}(i)_{[10]}$& 306& 326&331&258&229& 97&10&0&0 \\
\hline
$\overline{w}(i)_{[20]}$& 204&384&438&251&211& 74&5&0&0 \\
\hline
$\overline{w}(i)_{[30]}$& 282& 364& 413& 295& 177& 39&0&0& 0 \\
\hline
\end{tabular}
\end{center}
\caption{The empirical $\overline{w}(i)_{[m]}$
for $i=10-18$ and all maturities $[m]$.\label{t:w2}}
\end{table}


\begin{table} 
\vskip 0.5 true cm
\begin{center}
\begin{tabular}
{|c|c|c|c|c|c|c|c|c|c|c|c|c|}
\hline
$i_{\rm{bin}}$ &1&2&3&4&5&6&7&8   \\
\hline
$i\in i_{\rm{bin}}$ &1-3&4&5&6&7&8&9-10&11-17 \\
\hline
$\overline{w}(i_{\rm{bin}})_{[1]}$&1165&1269&1674&2147&1110&1147&1051&877 \\
\hline
\end{tabular}
\end{center}
\caption{Example of the empirical $\overline{w}(i_{\rm{bin}})$
for the maturity $[m]=[1]$.
The other $[m]$ can be computed from Table \re{t:w2}.
\label{t:wbin}}
\end{table}

\item{\bf{The empirical $\overline N_{\rm{data}}(v,i_{\rm{bin}})$ density
distributions}}

The final empirical density distribution $\overline N(v,i_{\rm{bin}})$ which
enters the $\chi$-squared merit
function is taken as
\be
\overline{N}_{\rm{data}}(v,i_{\rm{bin}})
       =\sum_{i\in i_{\rm{bin}}}\overline{w}(i)\,
\overline{N}(v,i)
\label{Idistexp}
\ee
\end{description}

\section{\bf{The definition of $N_{\rm{theory}}$ and the $\chi$-squared merit
function }\label{sec:theory}}

For each maturity,
the empirical distribution  $\overline N_{\rm{data}}(v,i)$ has to be compared to
the effective
distribution $\overline{N}_{\rm{theory}}(v,i)$ computed from the theoretical
distribution
evaluated at lag $L=1$ day, which means for $l=N_L$.
This distribution
$\overline{p}(N_L,v,i)$ is computed from the seed by $N_L$ successive
convolutions.

We have
\bea
\overline{N}_{\rm{theory}}(v,i)&=&\overline{w}(i)\, \overline{p}(N_L,v,i)
      \nonumber\\
\overline{N}_{\rm{theory}}(v,i_{\rm{bin}})&=&
      \sum_{i\in i_{\rm{bin}}}\overline{N}_{\rm{theory}}(v,i)   \ .
\label{Idistth}
\eea

\begin{description}
\item{\bf{The $\chi$-squared merit function}}

For each maturity $[m]$,
the following $\chi$-squared function is defined
\be
\chi^2_{[m]}=\sum_{i_{\rm{bin}}=1}^{8}\sum_{v=-300}^{300}
            \frac{(\overline{N}_{\rm{theory}}(v,i_{\rm{bin}})
            -\overline{N}_{\rm{data}}(v,i_{\rm{bin}}))^2}
            {\sigma(v,_{\rm{bin}})^2}   \ .
\label{chisquared}
\ee
In this formula, the numerator is the square of the difference between the
empirical
distribution \re{densitydisc} and the theoretical distribution \re{Idistth} for
a given maturity.. It
is a
measure
of the discrepancy between theory and experiment. The denominator is the square
of the modified error $\sigma(v,_{\rm{bin}})$
According to the previous discussion around \re{sigmamod}, this
$\sigma(v,_{\rm{bin}})$ is taken as
\bea
\sigma(v,i_{\rm{bin}})_{\rm{modified}}
   \!\!\!&=&\!\!\!\left(N_{\rm{data}}(v,i_{\rm{bin}})
   N_{\rm{theory}}(v,i_{\rm{bin}})\right)^{1/4}
   			\quad\!\!{\rm{if}}
			\ N_{\rm{data}}(v,i_{\rm{bin}})\neq 0
\nonumber\\
\sigma(v,i_{\rm{bin}})_{\rm{modified}}
   \!\!\!&=&\!\!\!\left(N_{\rm{theory}}(v,i_{\rm{bin}})\right)^{1/4}
   			\quad\quad\quad\quad\quad\quad {\rm{if}}
			\ N_{\rm{data}}(v,i_{\rm{bin}})= 0 \ .
\label{sigmamod2}
\eea

With our definitions, the bins where the number of events is zero do not
contribute to the $\chi$-squared and hence can be removed or kept without
altering the results of the minimisations.

Finally, let us note that the limits on the summation over $v$, $[-300,300]$,
have been chosen at the point where the distributions in $v$ have become
empirically completely negligible. For example, increasing
the limits from 300 to 400 basis points does not change the values of the
parameters provided by the minimization method in any appreciable way.

\end{description}

\section{Seed Parametrisation \label{sec:results}}

In preceding articles \cite{NP}, \cite{AN}, we have shown that the FRS
distributions can be fitted very closely by using Pad\'e Approximates $[0,4]$
whose
coefficients follow rather simple scaling
laws (see \re{p1dist}).

\subsection{The form of the Seed. Determination of the parameters
\label{sec:sleeper}}

Extrapolating the scaling laws discovered in \cite{NP}, \cite{AN}
for lags equal to an integer number of days
to values of the lag small compared to
one day, as is explained in Appendix \re{ap:seedform}, one is led to state
the following form for the discretized seed
\be
s_L\, {S}(vs_V,is_I)=\frac{1}{s_V}
\frac{\alpha(i)}
     {1+\gamma(i)\,\mid\! v\!\mid^{d(i)}}
     \quad{\rm{\ for\ }}v\neq 0 \ .
\label{eduseed}
\ee
It depends on three independent parameters $\alpha$, $\gamma$ and $d$ which
themselves depend on the initial interest rate $i$ and on the maturity
$[m]$.
In this equation, using our choices of $s_L$ \re{sL},
$s_V$ \re{sV} and $s_I$ \re{sI}, we see that $v$ and $i$ are integers,
that the variation of interest rate $V=s_V v$ has the dimension of basis point,
that the initial interest rate $I=s_I i$ has the dimension of percent,
that the contribution of $s_L S$ to the probability distribution has the
dimension of the inverse basis point
and that $\alpha$, $\beta$ and $d$ have no dimension.

It is first convenient to define $V_{transition}$ as the variation of interest
rate $V$
at the place where the transition between a constant behavior (in $v$)
and the power decrease of the seed occurs. More precisely, it is defined as the
value for which the second term in the
denominator of \re{eduseed} becomes equal to the first term in this denominator.
Thus when the second term becomes $1$.
It is linked to the parameter $\gamma$ by
\be
v_{transition}\approx\gamma^{-\frac{1}{d}}\ .
\label{vtransition}
\ee
Using the definition \re{Vgrid}, we see that this corresponds to a
value of the interest rate variation $V_{transition}$ expressed in basis point
of
\be
V_{transition}=v_{transition}\ s_V \ .
\ee

It is then easy to understand the meaning of the three parameters:

\begin{itemize}
\item
The parameter $\alpha$ is linked, in first approximation, to the constant
behaviour of the seed for (non-zero) variations of the interest rate
appreciatively smaller than the
transition value defined in \re{vtransition}.

\item
The parameter $\gamma$ is linked, in first approximation, to the
transition value of the variation of the interest rate \re{vtransition} between
a constant behaviour (for small $v$) and a power decrease
of the seed for large $v$.

\item
The parameter parameter $d$ is linked to the asymptotic power decrease of the
seed when the variation of the interest rate is large compared to
$v_{transition}$.

\end{itemize}

Guided by the findings of \cite{NP}, \cite{AN}, we have taken the following
dependence of the three parameters in terms of the initiail interest rate:
\bea
\alpha(i)&=&	\alpha _1 e^{\alpha_2 (i - i_0)}
	\nonumber\\
\gamma(i)&=&  \gamma_1 e^{\gamma_2 (i - i_0)}
	\nonumber\\
d(i)&=&	d _1 e^{d_2 (i - i_0)+d_3 (i - i_0)^2} \ .
\label{paracd}
\eea
Again $i$ is in percent and $i_0$ is arbitrarily but very conveniently chosen to
be 6.5 percents (see the discussion in Appendix \re{ap:seedform}). At this
point, there are thus seven parameters which are maturity dependent.

\subsection{Determination of the parameters
\label{sec:sleepam}}

To calibrate the model, we have then proceeded along the following steps.

\begin{itemize}
\item
We have first computed, for each maturity $[m]$, the values of the seven
parameters ($\alpha_1,\alpha_2,\gamma_1,\gamma_2,d_1,d_2,d_3$)
which minimize the corresponding $\chi^2_{[m]}$.

\item
Looking at these minimal values, we have discovered that
linear extrapolation in
$[m]$ are approximately at work. More precisely, the following parametrisation
with
straight lines is suggested
\bea
\alpha_1([m])&=&
              \left(\alpha_{11}+\alpha_{12}([m]-[m]_0)\right)10^{-5}
     \nonumber\\
\alpha_2([m])&=&
              \left(\alpha_{21}+\alpha_{22}([m]-[m]_0)\right)10^{-5}
     \nonumber\\
\gamma_1([m])&=&\left(\gamma_{11}+\gamma_{12}([m]-[m]_0)\right)10^{-6}
     \nonumber\\
\gamma_2([m])&=&\left(\gamma_{21}+\gamma_{22}([m]-[m]_0)\right)10^{-1}
     \nonumber\\
d_1([m])&=&\left(d_{11}+d_{12}([m]-[m]_0)\right)
     \nonumber\\
d_2([m])&=&   \left(d_{21}+d_{22}([m]-[m]_0)\right)10^{-3}
     \nonumber\\
d_3([m])&=&   \left(d_{31}+d_{32}([m]-[m]_0)\right)10^{-4} \ .
     \nonumber
\label{paramat}
\eea
The $[m]_0$ which appears in these formulae is arbitrarily chosen to
be $[15]$ (years) close to the middle of the interval
$\left[[m]=[1],[m]=[30]\right]$. The normalisation
$10^{-5},10^{-6},\dots$
have been chosen conveniently. Clearly, the parameters with a 2 as their second
index are related to
the slopes of the straight lines while the parameters with a 1 as their first
index are the ordinates of the lines for $[m]=[m]_0$.
Altogether, there are now fourteen constant parameters.

\item
To determine the empirical values of the fourteen parameters
we have chosen to minimize the total merit function
\be
\chi^2_{total}=\sum_{[m]\in \{[m]\}}\chi^2_{[m]}
\label{meritchi}
\ee
which is simply the sum of the $\chi$-squared for the eight maturities
\re{matset}.
See the discussion in Appendix \re{ap:seedform} where the
results are summarized in Tables \re{fourteen1}-\re{fourteen2}.

\item
It turns out that only eight of the fourteen parameters are really relevant and
that six of them, namely $\alpha_{21},\alpha_{22},d_{12},d_{21},d_{22},d_{31}$
can be safely put to zero (see the discussion in Appendix \re{ap:seedform})
\be
\alpha_{21}=\alpha_{21}
         =d_{12}=d_{21}=d_{22}=d_{31}=0    \ .
\label{zeroparam}
\ee

\item

A last minimisation, taking again $\chi^2_{total}$ as the merit function and
taking into account the restrictions \re{zeroparam}, has
produced the final values of the eight remaining
non-zero parameters as given in Table \re{nine1}.
\vskip 1 true cm
\begin{table} 
\vskip 0.5 true cm
\begin{center}
\begin{tabular}
{|c|c|c|c|c|c|c|c|c|}
\hline
  &$\alpha_{11}$ &$\alpha_{12}$
  &$\gamma_{11}$ &$\gamma_{12}$ &$\gamma_{21}$ &$\gamma_{22}$
  &$d_{11}$ &$d_{32}$
  \\
\hline
value &2.39 &0.018 &5.78 &0.20 &-2.90 &-0.100 &3.030 &0.909
  \\
\hline
C.I. &$\pm$ 0.08  &$\pm$ 0.004
      &$\pm$ 0.99  &$\pm$ 0.04    &$\pm$ 0.49    &$\pm$ 0.032
      &$\pm$ 0.024 &$\pm$ 0.067

  \\
\hline
\end{tabular}
\end{center}
\caption{The selected $\alpha$, $\gamma$, $d$ parameters
         with the choice $i_0=6.5$, $[m]_0=15$
         and $\alpha_{21}=\alpha_{21}
         =d_{12}=d_{21}=d_{22}=d_{31}=0$.
         The C.I. are the confidence intervals.
         \label{nine1}}
\end{table}
\vskip 1 true cm

\end{itemize}

\subsection{Comparison with the data. Discussion \label{sec:seeddata}}

We are now in a position for a detailed comparison of the results of our fits
with the data. The values obtained for the parameters are summarized in
Table \re{tablealpha} for $\alpha$, in Table \re{tablegamma} for $\gamma$, in
Table \re{tabled} for $d$ and in Table \re{vtrans} for $v_{transition}$.
They are given in natural units (see \cite{NP} for a precise discussion of these
units).

\begin{description}

\item{The parameter $\alpha$}

The parameter $\alpha$ does not depend of the initial interest rate and is a
slowly increasing function of the maturity $[m]$.  Its values (multiplied by
$10^5$) are expressed in inverse basis points and given in Table
\re{tablealpha}.

\begin{table} 
\vskip 0.5 true cm
\begin{center}
\begin{tabular}
{|c|c|c|c|c|c|c|c|c|c|}
\hline
[m] &[1]& [2]& [3]& [5]& [7]& [10]& [20]& [30] \\
\hline
$\alpha\times 10^{5}$& 2.14& 2.15& 2.17& 2.21& 2.24& 2.30& 2.4& 2.6 \\
\hline
\end{tabular}
\end{center}
\caption{The values of $\alpha\times 10^{5}$
as a function of $[m]$.\label{tablealpha}}
\end{table}

\vskip 1 true cm

\item{The parameter $\gamma$}

The parameter $\gamma$ depends both on the maturity and on the interest rate.
Its dimension is the inverse of the basis point raised to the power $1/d$.
Its is given in Table \re{tablegamma} multiplied by $10^6$. In fact
$v_{transition}$ defined before, directly derived from $\gamma$ and given in
Table \re{vtrans}, carries a more intuitive meaning.
\begin{table} 
\vskip 0.5 true cm
\begin{center}
\begin{tabular}
{|c|c|c|c|c|c|c|c|c|c|}
\hline
$\gamma\times 10^{6}$  & i=1 &i=3  &i=5& i=7 &  i=9  & i=11  & i=13  \\
\hline
[m]=[1]& 6.85& 5.08& 3.77& 2.8& 2.07 &1.54 &1.14 \\
\hline
[m]=[2]& 7.72& 5.61& 4.08& 2.96& 2.15 &1.57 &1.14 \\
\hline
[m]=[3]& 8.66& 6.17& 4.39& 3.13& 2.23 &1.59 &1.13 \\
\hline
[m]=[5]& 10.8& 7.38& 5.05& 3.46& 2.37 &1.62 &1.11 \\
\hline
[m]=[7]& 13.3& 8.74& 5.75& 3.78& 2.49 &1.63 &1.07 \\
\hline
[m]=[10]& 17.9& 11.1& 6.86& 4.25& 2.63 &1.63 &1.01 \\
\hline
[m]=[20]& 43.9& 22.3& 11.3& 5.71& 2.89 &1.46 &0.742 \\
\hline
[m]=[30]& 98.6& 40.9& 16.9& 7.02& 2.91 &1.2 &0.499 \\
\hline
\end{tabular}
\end{center}
\caption{The values of $\gamma\times 10^{6}$
as a function of $[m]$ and $i$.\label{tablegamma}}
\end{table}
\vskip 1 true cm

\item{The parameter $d$}

The parameter $d$ (which is a pure number) depends on the maturity and on the
initial rate. It is given
in Table \re{tabled}. It should be
remarked that it is empirically always close to 3. One may even wonder if the
exact value of 3 would not be a critical exponent of the problem and have
general validity.
\begin{table} 
\vskip 0.5 true cm
\begin{center}
\begin{tabular}
{|c|c|c|c|c|c|c|c|c|c|}
\hline
$d$  & i=1 &i=3  &i=5& i=7 &  i=9  & i=11  & i=13  \\
\hline
[m]=[1]& 2.92& 2.98& 3.02& 3.03& 3.01 &2.95 &2.87 \\
\hline
[m]=[2]& 2.92& 2.99& 3.02& 3.03& 3.01 &2.96 &2.88 \\
\hline
[m]=[3]& 2.93& 2.99& 3.02& 3.03& 3.01 &2.96 &2.89 \\
\hline
[m]=[5]& 2.95& 3.0& 3.02& 3.03& 3.01 &2.98 &2.92 \\
\hline
[m]=[7]& 2.96& 3.0& 3.03& 3.03& 3.02 &2.99 &2.94 \\
\hline
[m]=[10]& 2.99& 3.01& 3.03& 3.03& 3.02 &3.0 &2.97 \\
\hline
[m]=[20]& 3.07& 3.05& 3.03& 3.03& 3.04 &3.06 &3.09 \\
\hline
[m]=[30]& 3.16& 3.08& 3.04& 3.03& 3.06 &3.12 &3.21 \\
\hline
\end{tabular}
\end{center}
\caption{The values of $d$
as a function of $[m]$ and $i$.\label{tabled}}
\end{table}

\vskip 1 true cm

\item{The transition point $v_{transition}$}

With the values of the parameters in \re{nine1}, the values of the transition
points $v_{transition}$ are given in Table \re{vtrans} for the odd $i$'s and all
the maturities.
One sees that, as expected by intuitive arguments, $v_{transition}$  is an
increasing function
of the initial interest rate and a decreasing function of the maturity.

\begin{table}
\vskip 0.5 true cm
\begin{center}
\begin{tabular}
{|c|c|c|c|c|c|c|c|c|c|}
\hline
$v_{transition}$ & i=1 &i=3  &i=5& i=7 &  i=9  & i=11  & i=13  \\
\hline
[m]=[1]& 59& 60& 62& 68& 78 &93 &117 \\
\hline
[m]=[2]& 56& 57& 61& 67& 77 &92 &115 \\
\hline
[m]=[3]& 53& 55& 59& 66& 76 &90 &113 \\
\hline
[m]=[5]& 48& 52& 56& 64& 74 &88 &110 \\
\hline
[m]=[7]& 44& 48& 54& 62& 72 &87 &107 \\
\hline
[m]=[10]& 39& 44& 51& 59& 70 &85 &104 \\
\hline
[m]=[20]& 26& 34& 43& 54& 67 &81 &97 \\
\hline
[m]=[30]& 19& 27& 37& 50& 65 &80 &92 \\
\hline
\end{tabular}
\end{center}
\caption{The values of $v_{transition}$ (see \re{vtransition},\re{Vtransition}),
as a function of $[m]$ and $i$.
\label{vtrans}}
\end{table}

\vskip 1 true cm

\item{The $\chi$-squared for the best parameters}

The total $\chi$-squared as well as the $\chi$-squared for all maturities can be
found in table \re{chimat}. In general, as discussed in the
appendix, the reduced $\chi$-squared are close to one and hence very good,
except for $[m]=20$ where
there are some uncertainties in the data and to a lesser extend for $[m]=5$.
This is reflected in the  quality of the agreement between the empirical
distributions and the data as seen in the plots below.

\end{description}

\subsection{A few plots \label{sec:plots}}

All the plots, for lag one and higher, for all bins and for all maturities show
good to very good agreement between our
theoretical curves and the data provided by the FRS site \cite{FRS}. Both the
central region where the variation $v$ of the interest rates is around zero and
and the tails where $v$ becomes large are very well accounted for. We have
selected randomly
eleven plots \re{plot1}-\re{plot11} as examples of the fits. In particular, the
domains of the selected plots are given in detail in the figure captions.

\section{Conclusions\label{sec:conclu}}

In this paper, a microscopic theory of the term structure of interest rates has
been developed. Convolution techniques, implying about ten successive
convolutions, combined with time translation
invariance lead to a time scaled theory where the term
structure for practical lags (one day or more) can be deduced from a seed
function living at a relatively small lag scale (a few seconds).

The previously discovered scaling laws, which were found to be valid at lags
expressed in days, suggest forms for the possible seeds and imply a
discretisation of the problem.

Our new
results show that the scaling law assumptions are even simpler at the
microscopic lag scale.
Indeed, it is shown that the FRS data are amazingly well reproduced
(for a lag of one day or two days but the results easily extend to
higher lags)
by assuming that the seed has the critical form of a
self-organized \cite{BTW} econo-physical system . In other words, a very simple
power law
behavior emerges with essentially only one $v^{-d}$ term. The exponent $d$ is of
the order of three in close agreement with the tail behaviours obtained
previously using the Hill \cite{Hill} estimator.

These results open the door to two major issues, one rather theoretical and the
other
more practical and pragmatic.

Since the seed has such a simple scaling form, it suggests the existence of an
underlying statistical
model. The discovery of the basic ingredients and laws leading in a natural and
systematic way to the scaling
would be a major achievement which, if attained, may also lead to new
theoretical insights in related but
different contexts. We expect that a self-organized structure may be at work,
with $d=3$, or close to three, as critical exponent for all the maturities. We
hope to come back to
this issue.

Finally, one may wonder how this theory and its predictions can be efficiently
used in the
context of risk management (e.g. Value-at-Risk computation) and fixed-income
derivative
pricing. For the time being, studies are under way to measure the add-value of
using this model
instead of the traditional ones. First results will be provided in a near
future.

\vfill\eject

\appendix

\section{A na\"\i ve introduction to Dirac type
distributions \label{ap:delta}}

\subsection{Formal definition \label{sap:defdelta}}

A distribution $G(V)$ is a continuous linear functional on some space of
functions of a real variable $V$. The Dirac $\delta(V)$ distribution, which
has
compact support,
is defined on a continuous $C^{\infty}$ function $f(V)$ at $V=0$ by
\be
\left(\delta(V),f(V)\right)=f(0)   \ .
\label{deltadef}
\ee

The $j$th derivative $G^{[j]}$ of a distribution of compact support $G(V)$ is
defined
as
\be
\left(G^{[j]}(V),f(V)\right)=(-1)^j\left(G(V),f^{[j]}(V)\right)
\label{derivdist}
\ee
where
\be
f^{[j]}(V)=\frac{d^j f(V)}{dV^j}
\label{deriv}
\ee
is the $j$th derivative of the function $f(V)$.

In particular, the second derivative of the Dirac distribution
\be
\left(\delta^{[2]}(V),f(V)\right)=f^{[2]}(0)
\label{deltasecond}
\ee
is the second derivative of $f(V)$ evaluated at $V=0$.

Without being mathematically rigorous, these definitions can be seen in a
more
intuitive way. As Laurent Schwartz, the inventor of the general
distribution concept, used to caution : The formulae given below, with
physicist
notations, have to be used with
great care and have to be justified in a detailed and precise way. But he
also
recognized that they were very convenient to guess properties of
distributions.

\subsection{Discussion and approximations for the Dirac distributions\label{
sap:discDirac}}

The distribution $\delta(V)$ is
a generalized function
which can be thought as being zero for $V<0$ and $V>0$ and infinite at
$V=0$ in
such a way that, in formal agreement with \re{deltadef}
\be
\int_{-\infty}^{+\infty} \delta(V)\, f(V)\, dV=f(0) \ .
\label{deltatest}
\ee

There are many functions which can approximate the delta function. Let us
give
two.

\begin{enumerate}
\item
The step-type function
\be
\delta(V)=\lim_{s\rightarrow 0} D(V,s)
   \ \left\{
    \matrix{
    D(V,s)=0
              &\quad {\rm{for}}\quad
              & V<-\frac{s}{2}\cr
              &&					\cr
    D(V,s)=\frac{1}{s}
              &\quad {\rm{for}}\quad
              & -\frac{s}{2}\leq V\leq\frac{s}{2}\cr
              &&					\cr
    D(V,s)=0
              &\quad {\rm{for}}\quad
              & \frac{s}{2}< V
      }
   \right.
\label{stepdelta}
\ee
has compact support. It is not difficult do the Riemann integration and
prove
that for some $\hat V$
\be
-\frac{s}{2} \leq \hat V\leq \frac{s}{2}
\ee
one has
\bea
\int_{-\infty}^{+\infty} D(V,s)\, f(V)\, dV
&=&f(\hat V)
\left(\int_{-\frac{s}{2}}^{+\frac{s}{2}} D(V,s) \,dV\right)
      \nonumber\\
&=&f(\hat V)\ .
\label{deltanorm}
\eea
The limit for $s\rightarrow 0$ reproduces clearly \re{deltatest}
as $f(\hat V)\rightarrow f(0)$.

\item
The Dirac distribution can also be approximated by the continuous
function
\bea
\delta(V)&=&\lim_{\Delta\rightarrow 0}E(V,\Delta)
       \nonumber \\
E(V,\Delta)&=&\frac{\Delta}{\pi\left(\Delta^2+V^2\right)} \ .
\label{delta02}
\eea
This form is very interesting, as it shows that the $\delta(V)$ function is
the
boundary value of a continuous function of two variables. For positive
$\Delta$,
this function is purely positive for all $V$.

\end{enumerate}

A step-type approximation for the second derivative of the delta function is
given as follows. Consider the intervals
$A=[-3s/2,-s/2],B=[-s/2,+s/2],A=[+s/2,+3s/2]$, the step function which is zero
outside the three intervals, has value $1/s$ in the intervals $A$ and $C$ and
$-2/s$ in the interval $B$ and let $s\rightarrow 0$.

\section{Formal discussion of the Master Equation.
Continuous variables \label{ap:formal}}

In the following it is shown that the formal solution of the master equation is
fully compatible with the convolution equation \re{convolution} whatever
be the
intermediate time

Suppose that, the probability distribution has a Taylor power
expansion, around  $L=0$, of the form
\be
p(L,V,I_0)=\sum_{n=0}^{\infty}\ \frac{L^n}{n!}\ p^{(n)}(V,I_0) \ .
\label{powerexp}
\ee
where $p^{(n)}(V,I_0)$ is the $n$'th order derivative of $p$ with respect to
$L$
evaluated at $L=0$
\be
p^{(n)}(V,I_0)=\partial_{\lower0.5ex\hbox{\scriptsize{\it{L}}}}^n
		\, p(L,V,I_0)\,
        {\vrule height11pt width0.4pt depth 10pt}_{L=0} \ .
\label{derLn}
\ee
The zeroth order $p^{(0)}(V,I_0)$ is known by Eq.\re{deltaprob} and the
firth
order $p^{(1)}(V,I_0)$ by Eq.\re{seed}
\bea
p^{(0)}(V,I_0)&=&p(0,V,I_0)=\delta(V)
           \nonumber\\
p^{(1)}(V,I_0)&=&S(V,I_0)   \ .
\label{order01}
\eea
Introducing this power expansion in the master equation \re{master} and
equating, in the left and right hand sides, the coefficients of the same
powers
in $L$, one gets  the recurrence equations
\be
p^{(n)}(I_f-I_0,I_0)=\int_{-\infty}^{+\infty}
      p^{(n-1)}(I_f-I_i,I_i)\, S(I_i-I_0,I_0) \, dI_i
      \ ,\quad n=1,\ldots,\infty\ .
\label{recurrence}
\ee
It is easily seen that the equation for $n=1$ is an identity. For $n=2$, one
finds
\be
p^{(2)}(I_f-I_0,I_0)=\int_{-\infty}^{+\infty}
               S(I_f-I_1,I_1)\, S(I_1-I_0,I_0) \, dI_1
\label{order2}
\ee
where the dummy integration variable has been called $I_1$.
For arbitrary $n\geq 1$, one obtains
\bea
p^{(n+1)}(I_f-I_0,I_0)&=&
\int_{-\infty}^{+\infty}\, dI_{n} \int_{-\infty}^{+\infty}\, dI_{n-1}
            \ldots\int_{-\infty}^{+\infty}\, dI_1
      \nonumber\\
     &&\quad\quad\quad S(I_f-I_{n},I_n)\,S(I_n-I_{n-1})\ldots S(I_1-
     I_0,I_0)
      \nonumber\\
         &=&
\int_{-\infty}^{+\infty}\left(\prod_{p=1}^n \, dI_p\right)
             \left(\prod_{q=0}^{n} S(I_{q+1}-I_q,I_q)\right)
                  \ ,\ \ I_{n+1}=I_f \ .
\label{ordern}
\eea

It is not difficult to check that the formal power expansion
\re{powerexp}, \re{order01}, \re{order2}, \re{ordern} satisfies the
convolution
equation \re{convolution} without any further condition. Hence, it can
reasonably be supposed that one can focus  on the master
equation
\re{master} together with its natural boundary conditions.

\section{Master Equation with a simplified Scaling Law
and no $I_0$ dependence. The Gauss distribution solution
{\label{ap:gauss}}}

As suggested in section \re{sec:master2}, it is convenient to study analytically
the solution of the master equation with the seed \re{seeddelta}
\be
S(V,I_0)=\kappa(I_0)\,\partial_V^2\,\delta(V)\ .
\ee

In order to simplify the problem let us limit ourselves to suppose that the
distribution $p$ does not depend on $I_0$
\be
p(L,V,I_0)=\widehat{p}(L,V)
\label{pnoI0}
\ee
as well as the seed \re{seeddelta} and hence
\be
\kappa(I_0)=\widehat{\kappa}\ .
\label{kappanoI0}
\ee
Eq.\re{master2} becomes
\be
\partial_{\lower0.5ex\hbox{L}}\,\hat{p}(L,V)
      =\widehat{\kappa}\,
      \partial^2_{V}\,\widehat{p}(L,V)\ .
\label{master3}
\ee

To simplify the problem even further,
``scaling laws''
which are approximately but not exactly verified by the data \cite{AN}, are
assumed.
It has to be emphasized that the data do not support exactly these
scaling laws. Hence the results we will obtain in this section cannot be
hoped to be correct. On
the other hand, it is worth to quote them since they have as a consequence
the
very common but wrong belief that, in a natural way, distributions should
fall off at large $V$ as exponentials.

The ultra simplification of ¨the empirical results found in \cite{NP}, \cite{AN}
is
equivalent to the statement that the distributions depend on the reduced
variable $\wt{V}$ \re{mas4distb} and scales \re{mas4dista} in the natural way as
$1/\sqrt{L}$
\bea
p(L,V,I_0)&=&\frac{\wt{p}({\wt{V}})}{\sqrt{L}}
             \label{mas4dista}\\
     \wt{V}&=&\frac{V}{\sqrt{L}}  \ .
\label{mas4distb}
\eea

Replacing $p$ by its guess \re{mas4dista} in \re{master2}, one finds the final
equation
\be
2\kappa\,
\partial^2_{\lower0.5ex\hbox{{$\wt{V}$}}}\,\wt{p}({\wt{V}})
+\partial_{\lower0.5ex\hbox{{$\wt{V}$}}}\,
\biggl({\wt{V}}\,{\wt{p}}({\wt{V}})\biggr)
= 0    \ .
\label{master4}
\ee
The solutions of this equation can easily be found. Indeed, it can be
written
\be
\partial_{\lower0.5ex\hbox{{$\wt{V}$}}}\,
\biggl(2\kappa \partial_{\lower0.5ex\hbox{{$\wt{V}$}}}\,
				{\wt{p}}({\wt{V}})
                 +\wt{V}\,\wt{p}({\wt{V}})\biggr)=0
\label{master4b}
\ee
whose general solution is
\be
2\kappa \partial_{\lower0.5ex\hbox{{$\wt{V}$}}}\,
			{\wt{p}}({\wt{V}})
                 +\wt{V}\,\wt{p}({\wt{V}})= R_1
\label{master4c}
\ee
where $R_1$ is an arbitrary constant. To solve this
first order differential
equations, the homogeneous equation is solved
\be
2\kappa \partial_{{\lower0.5ex\hbox{$\wt{V}$}}}\,
			{\wt{p}^h}({\wt{V}})
                 +\wt{V}\,\wt{p}^h({\wt{V}})=0
\label{master4d}
\ee
giving the homogeneous solution $\wt{p}^h({\wt{V}})$
\be
\wt{p}^h({\wt{V}})=H\,e^{-\frac{\wt{V}^2}{4\kappa}}   \ .
\label{master4sol}
\ee
Assuming (the method of variation of constants) that $H$ depends
on
$\wt{V}$, introduce \re{master4sol} in \re{master4c} to obtain the
equation
\be
\partial_{\lower0.5ex\hbox{{$\wt{V}$}}}\,
	H(\wt{V})=\frac{R_1}{2\kappa}
		e^{\frac{\wt{V}^2}{4
\kappa}}
\ee
which can be solved by a simple integration
\be
H(\wt{V})=\frac{R_1}{2\kappa}
     \int_0^{\wt{V}}\,dx\,e^{\frac{x^2}{4\kappa}} +R_2
\label{Hsol}
\ee
where $R_2$ is the second constant of integration.
One has now to plot the final solution obtained by replacing the solution
for
$H$ \re{Hsol} in the homogeneous solution \re{master4sol}
\bea
\wt{p}({\wt{V}})&=&H(\wt{V})\,e^{-\frac{\wt{V}^2}{4\kappa}}
             \nonumber\\
             &=&\biggl(
       \frac{R_1}{2\kappa}\int_0^{\wt{V}}\,dx\,e^{\frac{x^2}{4\kappa}}
         +R_2\biggr)
      \,e^{-\frac{\wt{V}^2}{4\kappa}}   \ .
\label{master4solb}
\eea

In fact, we are looking for a distribution function $p(L,V,I_0)$ which is, in
first order symmetric in $V$, hence for a scaled function $\wt{p}(\wt{V})$
which
is also symmetrical in $\wt{V}$. Since by \re{master4sol} the
homogeneous
solution $\wt{p}^h(\wt{V})$ is symmetrical, one looks for a symmetrical
$H(\wt{V})$ solution. This clearly implies that $R_1$ should be zero.

We thus conclude that the solution of the master equation with strict
scaling law leads to the familiar Gauss type behavior
\be
\wt{p}({\wt{V}})=R_2\,e^{-\frac{\wt{V}^2}{4\kappa}}
\label{master4solc}
\ee
which is not at all sustained by the data as the main result of our preceding
investigation has shown that the asymptotically the term structure decreases as
a
power
of $1/V^d$ with $d$ around three to four.

\section{Form of the Seed. Phenomenological Discussion\label{ap:seedform}}

In preceding articles \cite{NP}, \cite{AN}, distributions fitting very
closely
the FRS data have been obtained using Pad\'e Approximants $[0,4]$
(see \re{p1dist}) i.e. a
polynomial of zero degree in $v$ in the numerator divided by a polynomial of
fourth degree in the denominator
\be
p(v,i)=
\frac{a(i)}{1+b(i)v^2+c(i)v^4} \ .
\label{p1dist}
\ee
Moreover, it was shown that the Pad\'e
coefficients ($a(i),b(i),c(i)$), which also depend on the maturity, follow
rather simple scaling laws.
Extrapolating these scaling laws for values of the lag small compared to
one day have lead us to guess suitable forms for the seed.
Though, the asymptotic behavior $\mid\!v\!\mid^{-d}$ for large $v$ with $d$
equal four which follows from \re{p1dist} is not incompatible with the data,
Hill estimators \cite{Hill} were
pointing
towards a somewhat smaller value of $d$. Precise values can be found in
\cite{NP}.

Guided by this work and by general ideas about scaling laws and self-organized
criticality, we have progressively been led to a simpler educated guesses for
the seed, namely \re{eduseed},
where three parameters $\alpha$, $\gamma$ and $d$, in principle, depend on the
initial interest rate $i$ and
maturity $[m]$. For the dependence in $i$, we were led to
\re{paracd} where $i_0$ can be arbitrarily chosen.
The seven coefficients themselves are thought to be, in first approximation
linear in the maturity $m$. More precisely, we have thus used
\re{paracd} and \re{paramat} with their
convenient normalisations. They depend on
initially on fourteen independent constant parameters.

At first, we have chosen, as is generally advisable to diminish the
relationships
between the relevant parameters, the arbitrary $i_0$ an $[m]_0$ in the middle
of their respective domains and more precisely $i_0=6$ and $[m]=[15]$.
Starting with this form of the seed and allowing for $N_L=10$ iterations (see
the discussion
following eq.\re{Lgrid}, after minimizing the total $\chi$-squared we find the
results of Tables \re{fourteen1} and \re{fourteen2} for the fourteen parameters
in
\re{paramat}.
The values of the resulting $\chi^2_{[m]}$ and $\chi^2_{total}$ are given in
Table \re{chimat} together with the number of degrees of freedom $N_f$ and the
reduced $\chi$-squared $\chi^2/N_f$
for each maturity.

\begin{table} 
\vskip 0.5 true cm
\begin{center}
\begin{tabular}
{|c|c|c|c|c|c|c|c|c|}
\hline
  &$\alpha_{11}$ &$\alpha_{12}$ &$\alpha_{21}$ &$\alpha_{22}$
  &$\gamma_{11}$ &$\gamma_{12}$ &$\gamma_{21}$ &$\gamma_{22}$
  \\
\hline
value &2.46   &0.022   &0.005   &-0.015
      &6.53   &0.230   &-2.90   &-0.119
  \\
\hline
C.I. &$\pm$ 0.08  &$\pm$ 0.0042   &$\pm$ 1.57    &$\pm$ 0.122
      &$\pm$ 0.99  &$\pm$ 0.039    &$\pm$ 0.49    &$\pm$ 0.032
  \\
\hline
\end{tabular}
\end{center}
\caption{The $\alpha$ and $\gamma$ parameters
for the initial arbitrary choice of $i_0=6$ and $[m]_0=[15]$.
The C.I. are the confidence intervals.\label{fourteen1}}
\end{table}

\begin{table} 
\vskip 0.5 true cm
\begin{center}
\begin{tabular}
{|c|c|c|c|c|c|c|}
\hline
  &$d_{11}$ &$d_{12}$ &$d_{21}$ &$d_{22}$ &$d_{31}$ &$d_{32}$
  \\
\hline
value &3.015  &0.0001  &0.04   &0.005
      &-0.01 &0.907
  \\
\hline
C.I. &$\pm$ 0.024 &$\pm$ 0.0005   &$\pm$ 2.47    &$\pm$ 0.161
      &$\pm$ 0.82  &$\pm$ 0.067
  \\
\hline
\end{tabular}
\end{center}
\caption{The $d$ parameters
for the initial arbitrary choice of $i_0=6$ and $[m]_0=[15]$ \label{fourteen2}}.
The C.I. are the confidence intervals.
\end{table}

\begin{table} 
\vskip 0.5 true cm
\begin{center}
\begin{tabular}
{|c|c|c|c|c|c|c|c|c|c|}
\hline
  &$[1]$ &$[2]$ &$[3]$ &$[5]$ &$[7]$ &$[10]$ &$[20]$ &$[30]$ &$total$
  \\
\hline
$\chi^2$             &1064  &1164  &930   &1440
                     &1090  &996   &1832  &828   &9347.44
  \\
\hline
$N_{f}$              &1099  &931   &959   &871
                     &853   &793   &691   &651   &6911
  \\
\hline
$\chi^2/N_f$         &0.97  &1.25  &0.97  &1.65
                     &1.28  &1.26  &2.65  &1.27  &1.35
  \\
\hline
\end{tabular}
\end{center}
\caption{The $\chi^2$, the reduced $\chi^2/N_{f}$  and the number of degree of
freedom $N_{f}$
\label{chimat}}
\end{table}

Let us now discuss the results of the Tables \re{chimat}, \re{fourteen1},
\re{fourteen2} and their implications for the
choice of the parameters.
\begin{description}
\item{The $\chi^2$}

From Table \re{chimat} we see that the reduced $\chi$-squared are rather
good except for $[m]=[20]$.
Let us recall that there is some incoherences for this
maturity as the data is composed of two disjointed sets of points corresponding
to two disjoined periods in time.

\item{The parameters $\alpha$}

From Table \re{fourteen1},
we see that two parameters $\alpha_{21}$ and $\alpha_{22}$ are very close to
zero with errors much larger than their values. This obviously points to a
zero value for the combined parameter $\alpha_2$ which, we recall, sets the
dependence in $i$ of the parameter $\alpha$. It means that whatever be the level
of the
interest rate, for very short intervals the probability of the occurrence of a
certain small variation (before the decrease in $|v|^{-d}$ plays a role) does
not depend on the interest rate. This is an interesting conclusion.
From Table \re{fourteen1},
we also learn that the two other parameters ($\alpha_{11}$ and
$\alpha_{12}$) are meaningful. We conclude that the parameter $\alpha$, which
essentially does
not depend on the initial interest rate, increases appreciably with the
maturity.

\item{The parameters $\gamma$}

The four $\gamma$ parameters are all significant. This means that
$v_{transition}$ (\re{vtransition} where the constant behaviour of the seed
tends to become a
power behaviour is highly dependent on the maturity and on the initial interest
rate. As discussed in the main part of the text, this behaviour conforms to our
expectations.

\item{The parameters $d$}

From Table \re{fourteen2},
we see that of the six $d$ parameters, four, namely
$d_{12},d_{22},d_{22},d_{31}$, can safely be put to zero. This means that $d$ is
almost constant for every maturity and every initial interest rate. Except for
$d_{11}$ which as expected is very close to 3, the only parameter which survive
is $d_{32}$. It follows that $d$ has a joint behaviour linear in maturity and
parabolic in the initial interest rate. Amazingly enough the arbitrary
parameters $i_0=6$ and $[m]_0=[15]$ which were chosen in the middle of their
respective domains were almost optimal choices which are further discussed
below.

\end{description}

In order to obtain the best value of $i_0$ leading to the best adjusted parabola
for the dependence of $d$ in $i$, we have rerun our minimization programs
putting to zero the six un-significant parameters
$\alpha_{21},\alpha_{22},d_{12},d_{21},d_{22},d_{31}$ but allowing the best
choice for the arbitrary parameter $i_0$. The results are given in
Table \re{nine1}. They show that $i_0$ around 6.5 would have even be a better
choice to
make $d_{12}$, $d_{21}$, $d_{22}$, $d_{31}$ equal to zero. In the main part of
the text we have
adhered to this choice.

\vskip .7 true cm
\noindent{\large\bf{Acknowledgment}}
\vskip 0.2 true cm
\noindent

The work of  Jean Nuyts was supported by the Belgian F.N.R.S. (Fonds National de
la Recherche Scientifique)
and the work of Thomas Alderweireld partly by the Belgian I.I.S.N (Institut
Inter-universitaire des Sciences Nucl\'eaires)
and the DEXIA group.

\newpage
\centerline{\bf{Figure Caption}}

\begin{description}
\refstepcounter{fig}

\vskip 0.5 true cm
\item{Figure \thefig.}
{\label{plot1}}
\refstepcounter{fig}
The comparison between the theoretical curve (line) and
the empirical curve (crosses) for $[m]=[1]$ and for the first bin
$i_{bin}=1$ which contains the initial
interest rates $i=1,2,3$. The related total number of events is found in
\re{t:wbin}. The horizontal axis is in basis points. The vertical
axis represents the number of events. The size of the arms of the crosses are
estimated errors equal approximatively to the square root of the number of
events. The curves for the bins $i_{bin}=4$ and $i_{bin}=8$
and for the same maturity $[m]=[1]$ are given
in Figures \re{plot2} and \re{plot3} respectively.
The analogous curves
for the other bins present fits of equivalent quality.

\vskip 0.5 true cm
\item{Figure \thefig.}
{\label{plot2}}
\refstepcounter{fig}

The comparison between the theoretical curve (line) and
the empirical curve (crosses) for $[m]=[1]$ and for the bin
$i_{bin}=4$ which contains the initial
interest rate $i=6$. The related total number of events is found in
\re{t:wbin}. The horizontal axis is in basis points. The
vertical axis represents the number of events. The size of the arms of the
crosses are estimated errors equal approximatively to the square root of the
number of events.
The curves for the bins $i_{bin}=1$ and $i_{bin}=8$
and for the same maturity $[m]=[1]$ are given
in Figures \re{plot1} and \re{plot3} respectively.
The analogous curves
for the other bins present fits of equivalent quality.

\vskip 0.5 true cm
\item{Figure \thefig.}
{\label{plot3}}
\refstepcounter{fig}

The comparison between the theoretical curve (line) and
the empirical curve (crosses) for $[m]=[1]$ and for the bin
$i_{bin}=8$  which contains the initial
interest rates $i=11-17$. The related total number of events is found in
\re{t:wbin}. The horizontal axis is in basis points. The
vertical axis represents the number of events. The size of the arms of the
crosses are estimated errors equal approximatively to the square root of the
number of events.
The curves for the bins $i_{bin}=1$ and $i_{bin}=4$
and for the same maturity $[m]=[1]$ are given
in Figures \re{plot1} and \re{plot2} respectively.
The analogous curves
for the other bins present fits of equivalent quality.

\vskip 0.5 true cm
\item{Figure \thefig.}
{\label{plot4}}
\refstepcounter{fig}

The comparison between the theoretical curve (line) and
the empirical curve (crosses)  for $[m]=[3]$ and
for the bin $i_{bin}=3$ which contains the initial
interest rate $i=5$. The related total number of events can be read in Table
\re{t:w1}. The horizontal axis is in basis points. The
vertical axis represents the number of events. The size of the arms of the
crosses are estimated errors equal approximatively to the square root of the
number of events.
The analogous curves for the same maturity $[m]=[3]$
for the other bins, as well as those for maturity $[m]=[2]$, present fits of
equivalent quality.

\vskip 0.5 true cm
\item{Figure \thefig.}
{\label{plot5}}
\refstepcounter{fig}

The comparison between the theoretical curve (line) compared with
the empirical curve (crosses)  for $[m]=[5]$ and
for the first bin $i_{bin}=4$ which contains the initial
interest rate $i=6$. The related total number of events can be read in
\re{t:w1}. The horizontal axis is in basis points. The
vertical axis represents the number of events. The size of the arms of the
crosses are estimated errors equal approximatively to the square root of the
number of events.
The analogous curves for the same maturity $[m]=[5]$
for the other bins, as well as those for maturity $[m]=[7]$, present fits of
equivalent quality.

\vskip 0.5 true cm
\item{Figure \thefig.}
{\label{plot6}}
\refstepcounter{fig}

The comparison between the theoretical curve (line) compared with
the empirical curve (crosses)  for $[m]=[10]$ and
for the first bin $i_{bin}=5$ which contains the initial
interest rate $i=7$. The related total number of events is found in
\re{t:w1}. The horizontal axis is in basis points. The
vertical axis represents the number of events. The size of the arms of the
crosses are estimated errors equal approximatively to the square root of the
number of events.
The analogous curves for the same maturity $[m]=[10]$
for the other bins present fits of equivalent quality.

\vskip 0.5 true cm
\item{Figure \thefig.}
{\label{plot7}}
\refstepcounter{fig}

The comparison between the theoretical curve (line) compared with
the empirical curve (crosses)  for $[m]=[20]$ and
for the first bin $i_{bin}=3$ which contains the initial
interest rate $i=5$. The related total number of events is found in
\re{t:w1}. The horizontal axis is in basis points. The vertical
axis represents the number of events. The size of the arms of the crosses are
estimated errors equal approximatively to the square root of the number of
events.
The analogous curves for the same maturity $[m]=[20]$
for the other bins present fits of equivalent quality.

\vskip 0.5 true cm
\item{Figure \thefig.}
{\label{plot8}}
\refstepcounter{fig}

The comparison between the theoretical curve (line) compared with
the empirical curve (crosses)  for $[m]=[30]$ and
for the first bin $i_{bin}=7$ which contains the initial
interest rates $i=9-10$. The related total number of events is found in
\re{t:w1}, \re{t:w2}. The horizontal axis is in basis points. The vertical
axis represents the number of events. The size of the arms of the crosses are
estimated errors equal approximatively to the square root of the number of
events.
The analogous curves for the same maturity $[m]=[30]$
for the other bins present fits of equivalent quality.

\vskip 0.5 true cm
\item{Figure \thefig.}
{\label{plot9}}
\refstepcounter{fig}

The comparison between the theoretical curve (line) compared with
the empirical curve (crosses) for the tail (i.e. for $v$ greater than 5 basis
points) for $[m]=[1]$
for the bin $i_{bin}=7$ which contains the initial interest rate $i=9-10$.
The horizontal axis is in basis points. The vertical axis represents the number
of events. The size of the arms of the crosses are estimated errors equal
approximatively to the square root of the number of events. The analogous curves
for the other bins and the other maturities present fits of equivalent quality.
The analogous curves for essentially all maturities
all bins, when they make sense, present fits of equivalent quality.

\vskip 0.5 true cm
\item{Figure \thefig.}
{\label{plot10}}
\refstepcounter{fig}

The comparison between the theoretical curve for $[m]=[5]$ and
for a Lag of 2 days (line) compared with
the empirical curve (crosses) for the first bin
$i_{bin}=3$ which contains the initial
interest rates $i=5$. The related total number of events is found in
\re{t:w2}. The horizontal axis is in basis points. The vertical
axis represents the number of events. The size of the arms of the crosses are
estimated errors equal approximatively to the square root of the number of
events.
We have checked that the analogous curves obtained by further convolutions,
for all maturities present
fits of equivalent quality.

\vskip 0.5 true cm
\item{Figure \thefig.}
{\label{plot11}}
\refstepcounter{fig}

The comparison between the theoretical curve for $]m]=[1]$ for a Lag of 2 days
(line) compared with
the empirical curve (crosses) for the tail (i.e. for $v$ greater than 5 basis
points) for the bin $i_{bin}=5$
which contains the initial interest rates $i=7$.
The horizontal axis is in basis points. The vertical axis represents the number
of events. The size of the arms of the crosses are estimated errors equal
approximatively to the square root of the number of events. The analogous curves
of the other bins present fits of equivalent quality.

\end{description}

\newpage

\centerline{\bf{Figure 1}}
\vskip 0.3 true cm
\centerline{Maturity=$[1]$, Initial Interest Rate $\approx 1{-}3$, Lag=1}
\begin{center}
\epsfxsize=10cm\epsffile{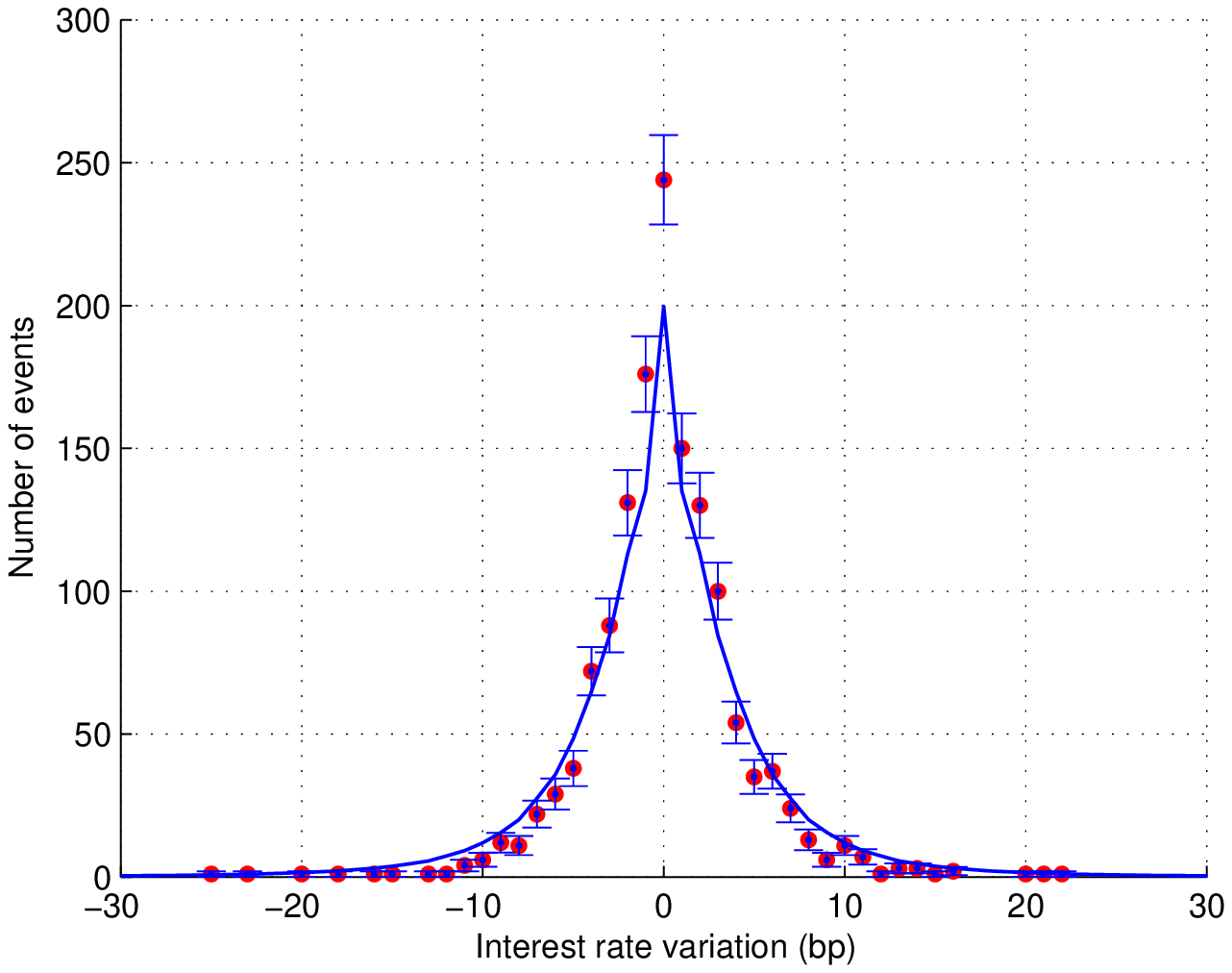}
\end{center}


\centerline{\bf{Figure 2}}
\vskip 0.3 true cm
\centerline{Maturity=$[1]$, Initial Interest Rate $\approx 6$, Lag=1}
\begin{center}
\epsfxsize=10cm\epsffile{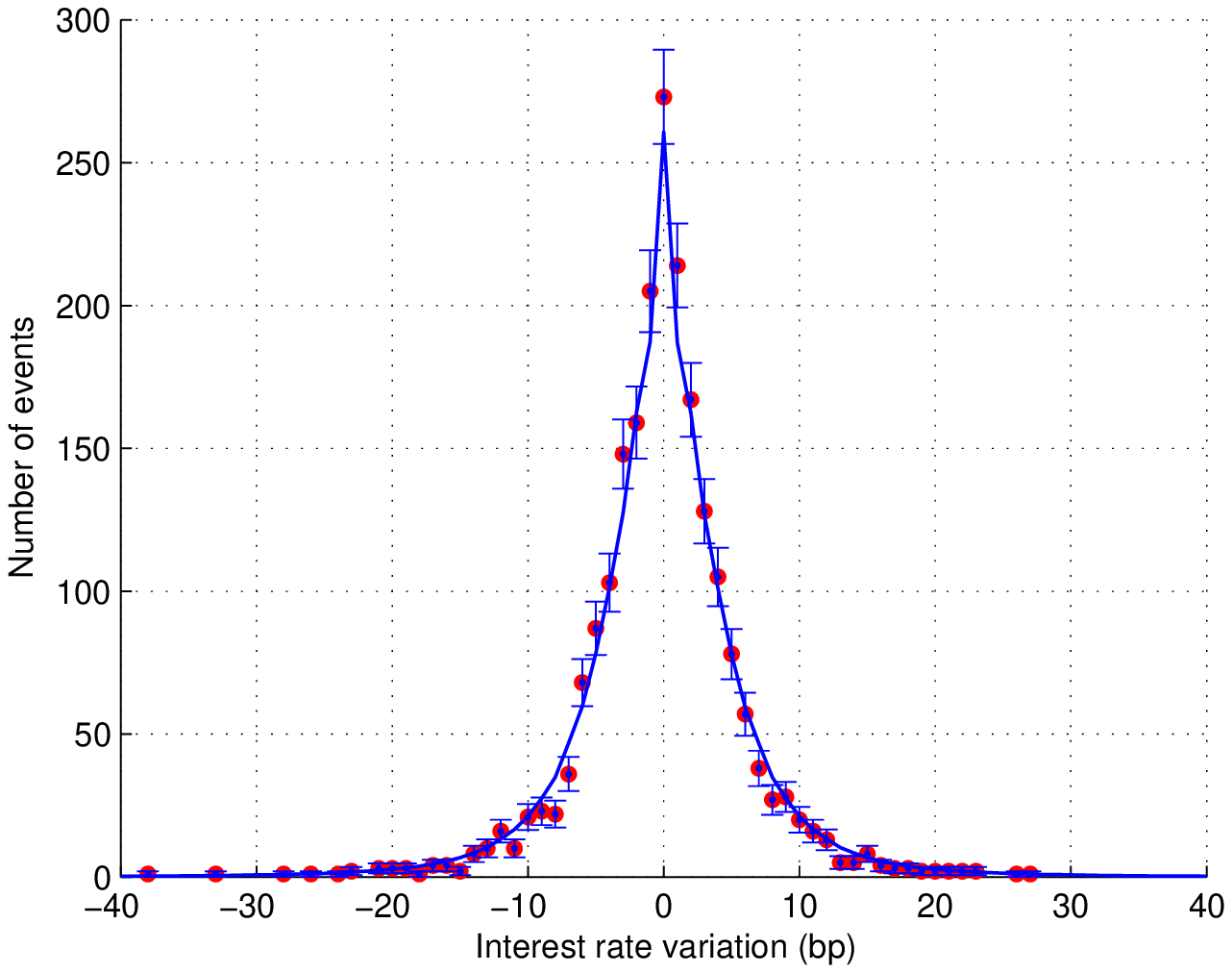}
\end{center}

\newpage

\centerline{\bf{Figure 3}}
\vskip 0.3 true cm
\centerline{Maturity=$[1]$, Initial Interest Rates $\approx 11{-}17$, Lag=1}
\begin{center}
\epsfxsize=10cm\epsffile{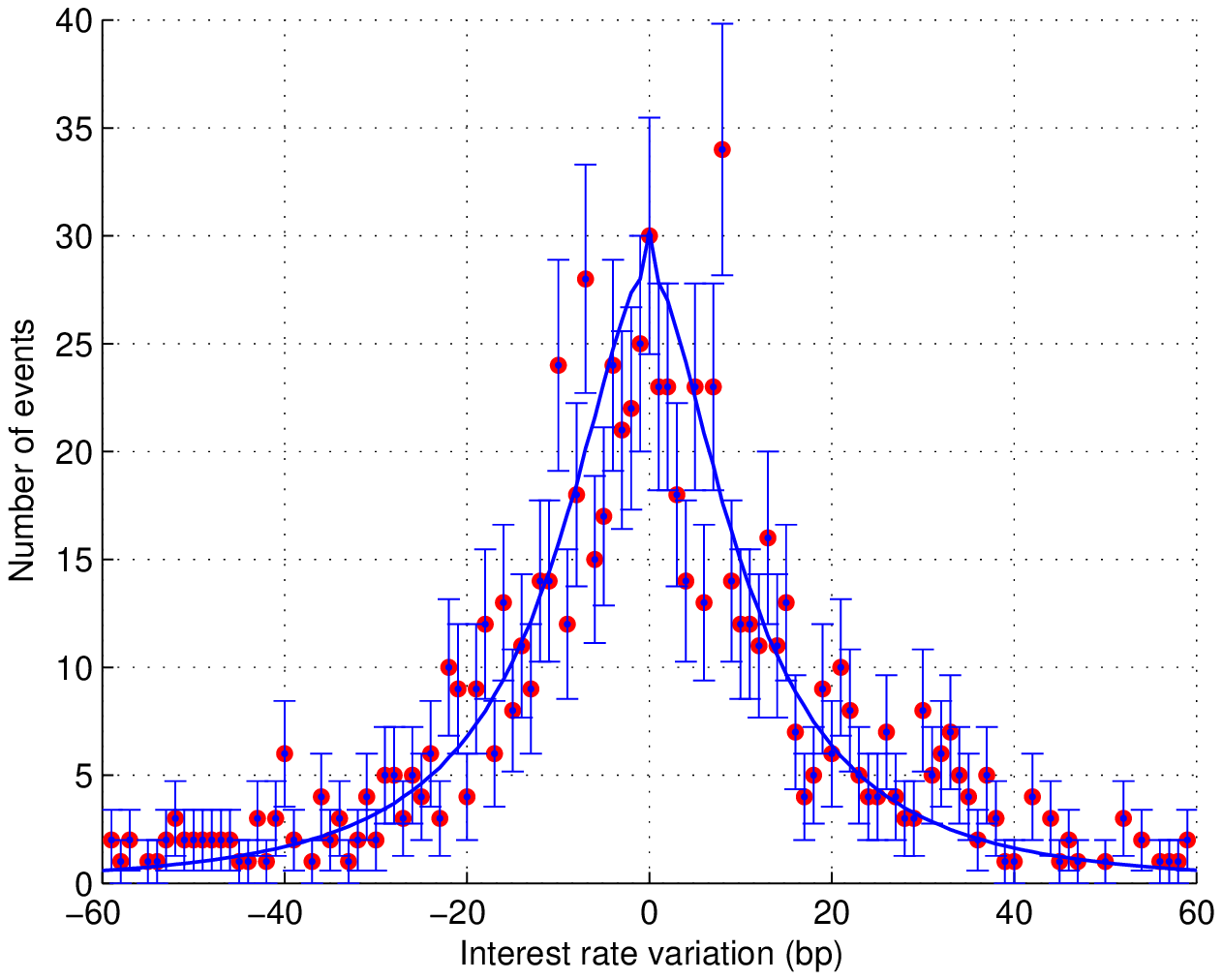}
\end{center}


\centerline{\bf{Figure 4}}
\vskip 0.3 true cm
\centerline{Maturity=$[3]$, Initial Interest Rate $\approx 5$, Lag=1}
\begin{center}
\epsfxsize=10cm\epsffile{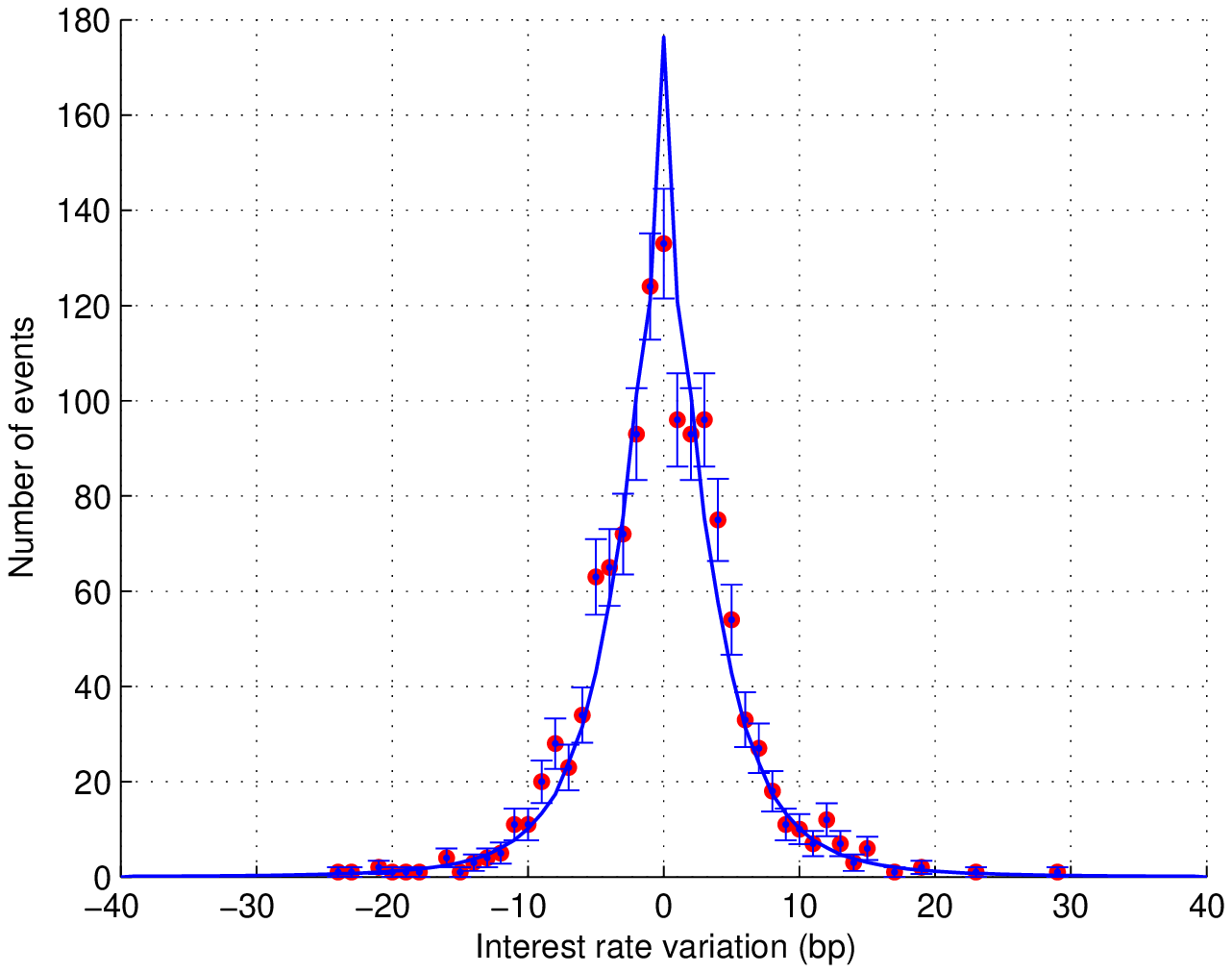}
\end{center}

\newpage

\centerline{\bf{Figure 5}}
\vskip 0.3 true cm
\centerline{Maturity=$[5]$, Initial Interest Rate $\approx 6$, Lag=1}
\begin{center}
\epsfxsize=10cm\epsffile{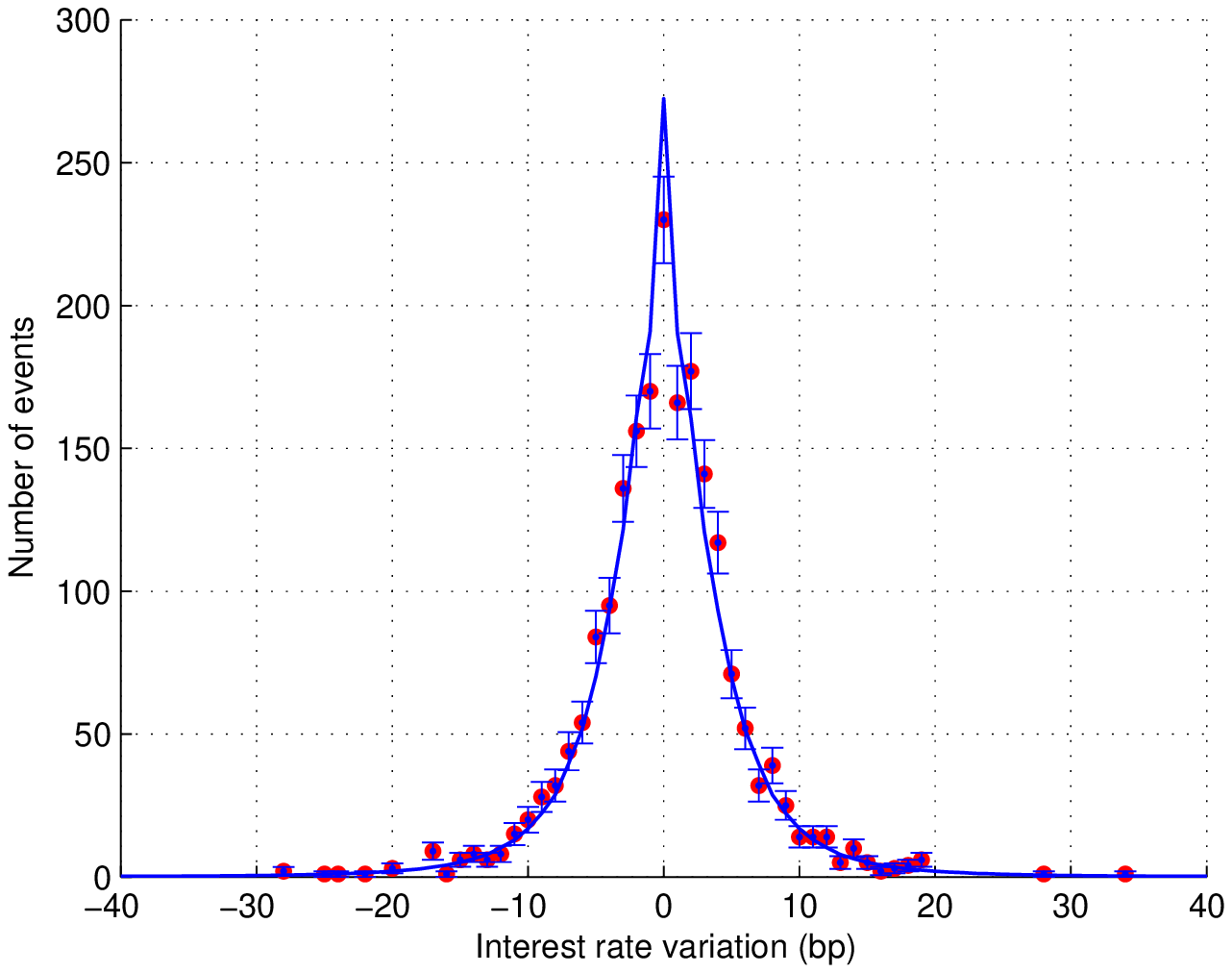}
\end{center}


\centerline{\bf{Figure 6}}
\vskip 0.3 true cm
\centerline{Maturity=$[10]$, Initial Interest Rate $\approx 7$, Lag=1}
\begin{center}
\epsfxsize=10cm\epsffile{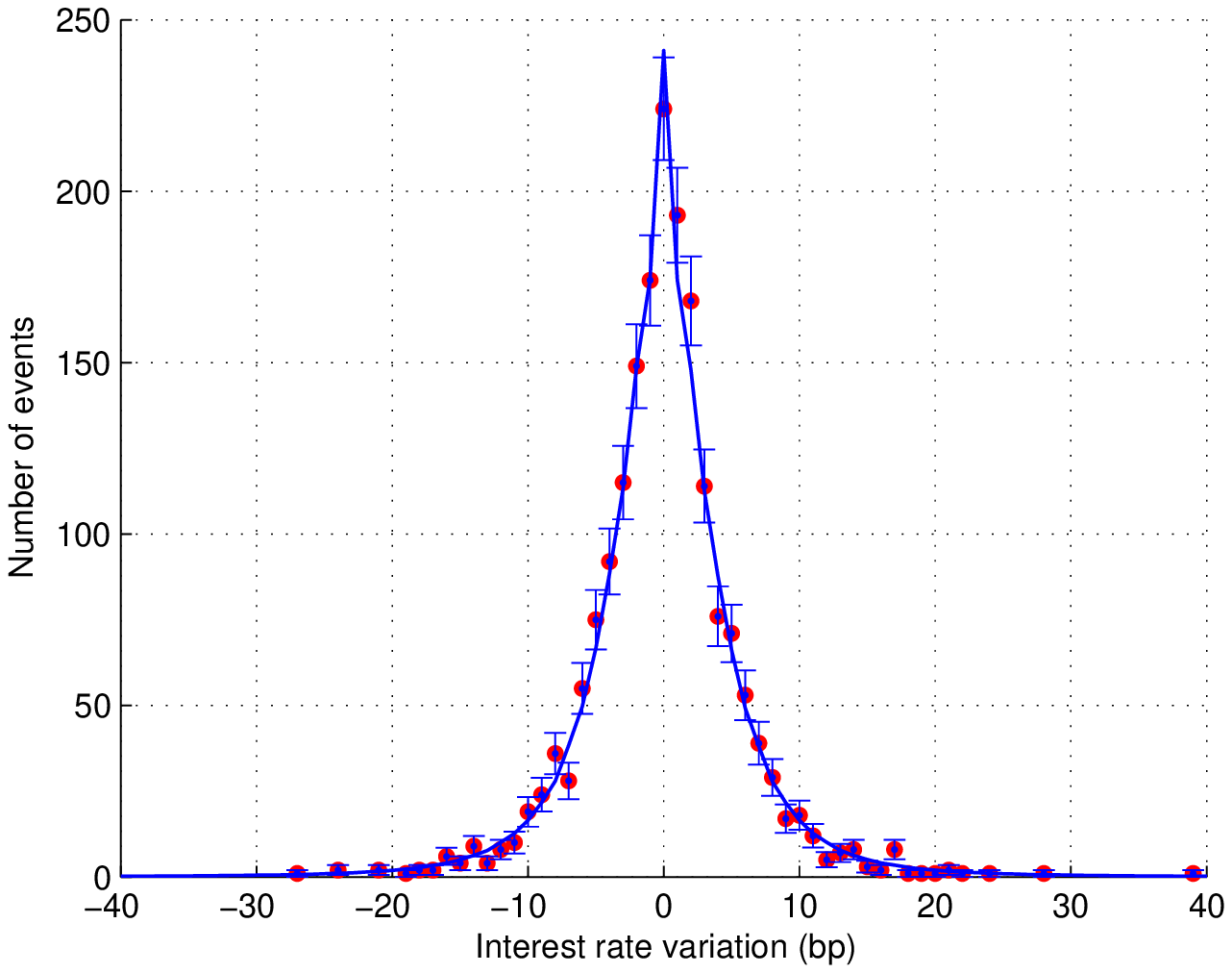}
\end{center}

\newpage

\centerline{\bf{Figure 7}}
\vskip 0.3 true cm
\centerline{Maturity=$[20]$, Initial Interest Rate $\approx 5$, Lag=1}
\begin{center}
\epsfxsize=10cm\epsffile{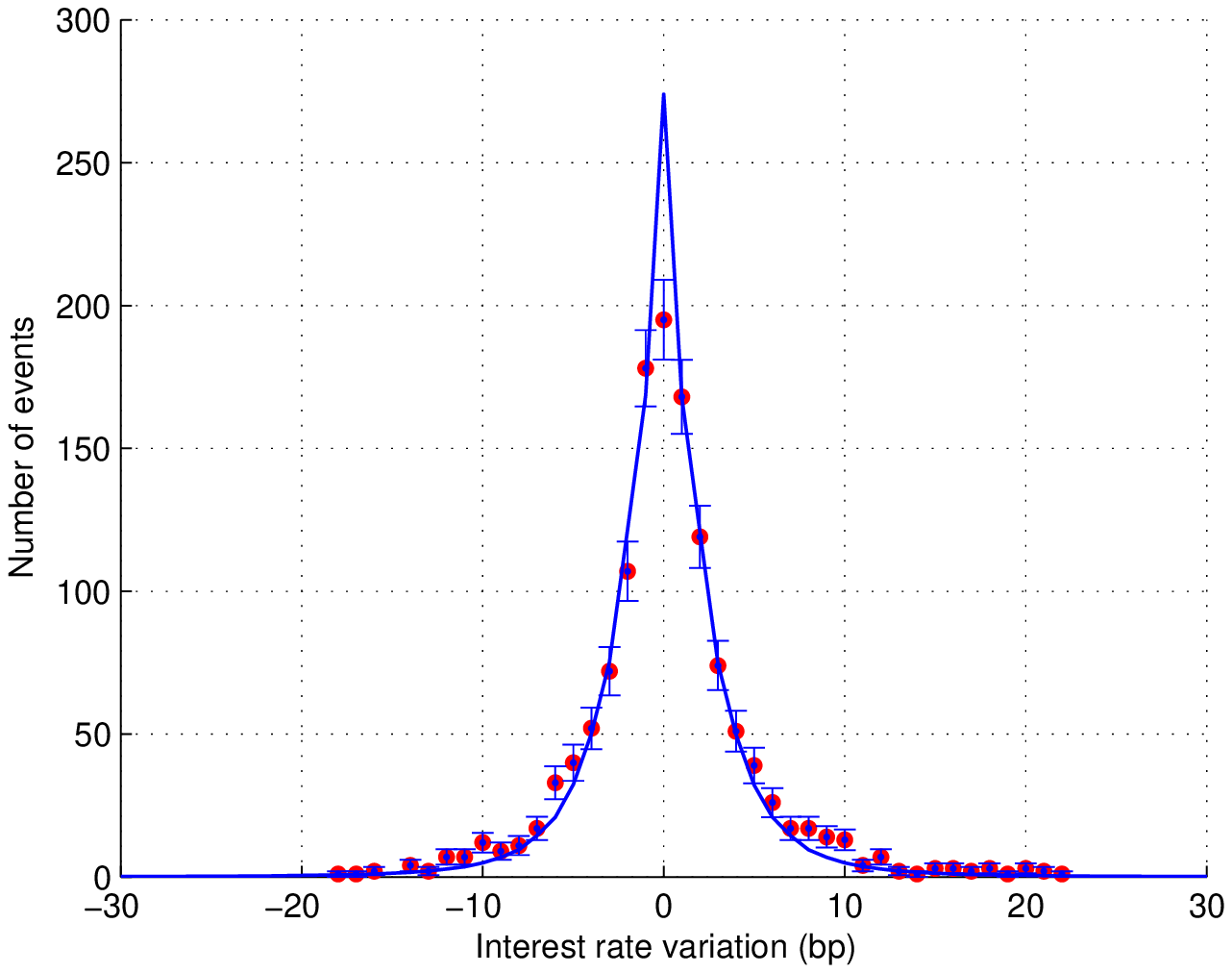}
\end{center}


\centerline{\bf{Figure 8}}
\vskip 0.3 true cm
\centerline{Maturity=$[30]$, Initial Interest Rates $\approx 9{-}10$, Lag=1}
\begin{center}
\epsfxsize=10cm\epsffile{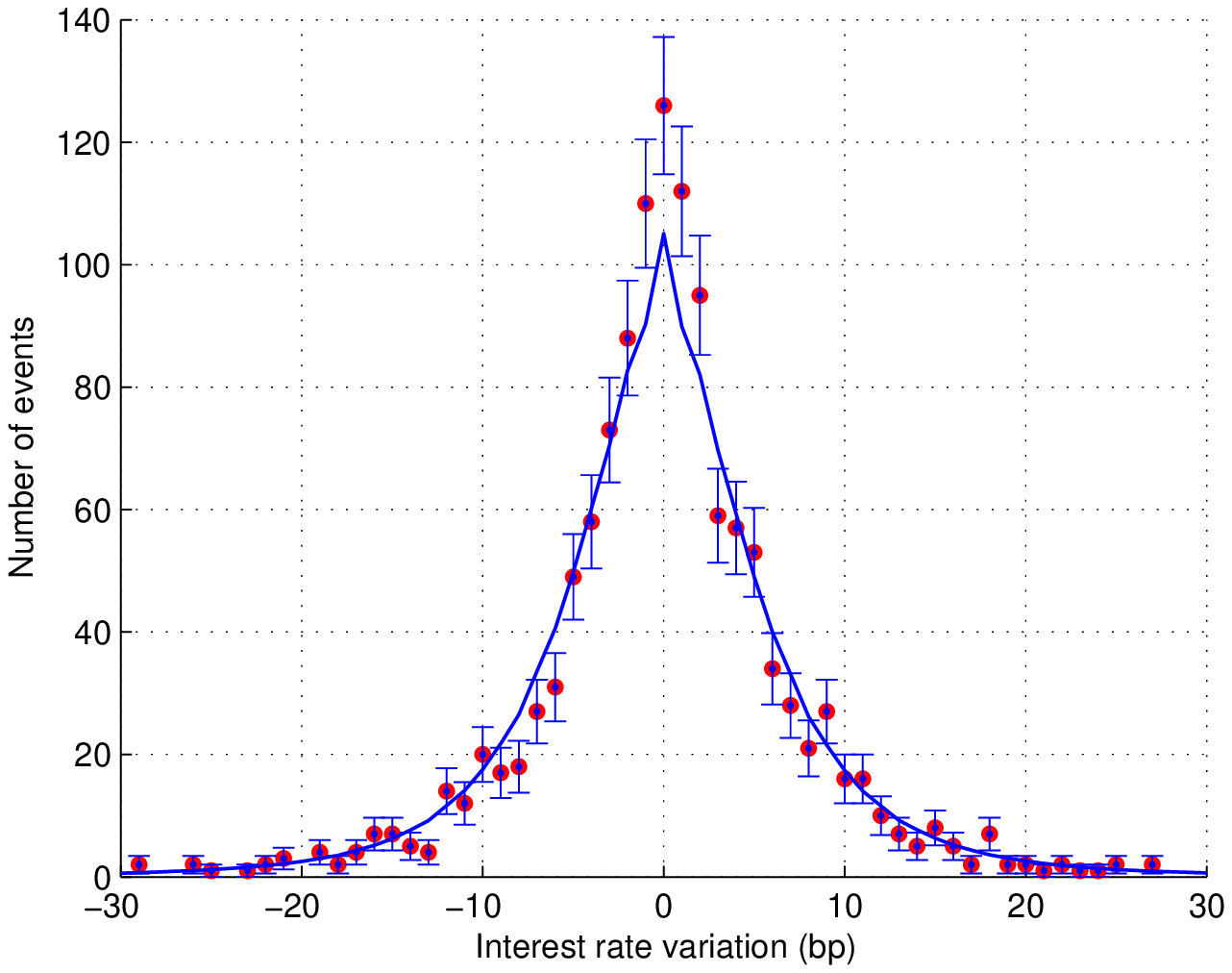}
\end{center}

\newpage

\centerline{\bf{Figure 9}}
\vskip 0.3 true cm
\centerline{Maturity=$[1]$, Initial Interest Rates $\approx 9{-}10$, Lag=1,
Tail}
\begin{center}
\epsfxsize=10cm\epsffile{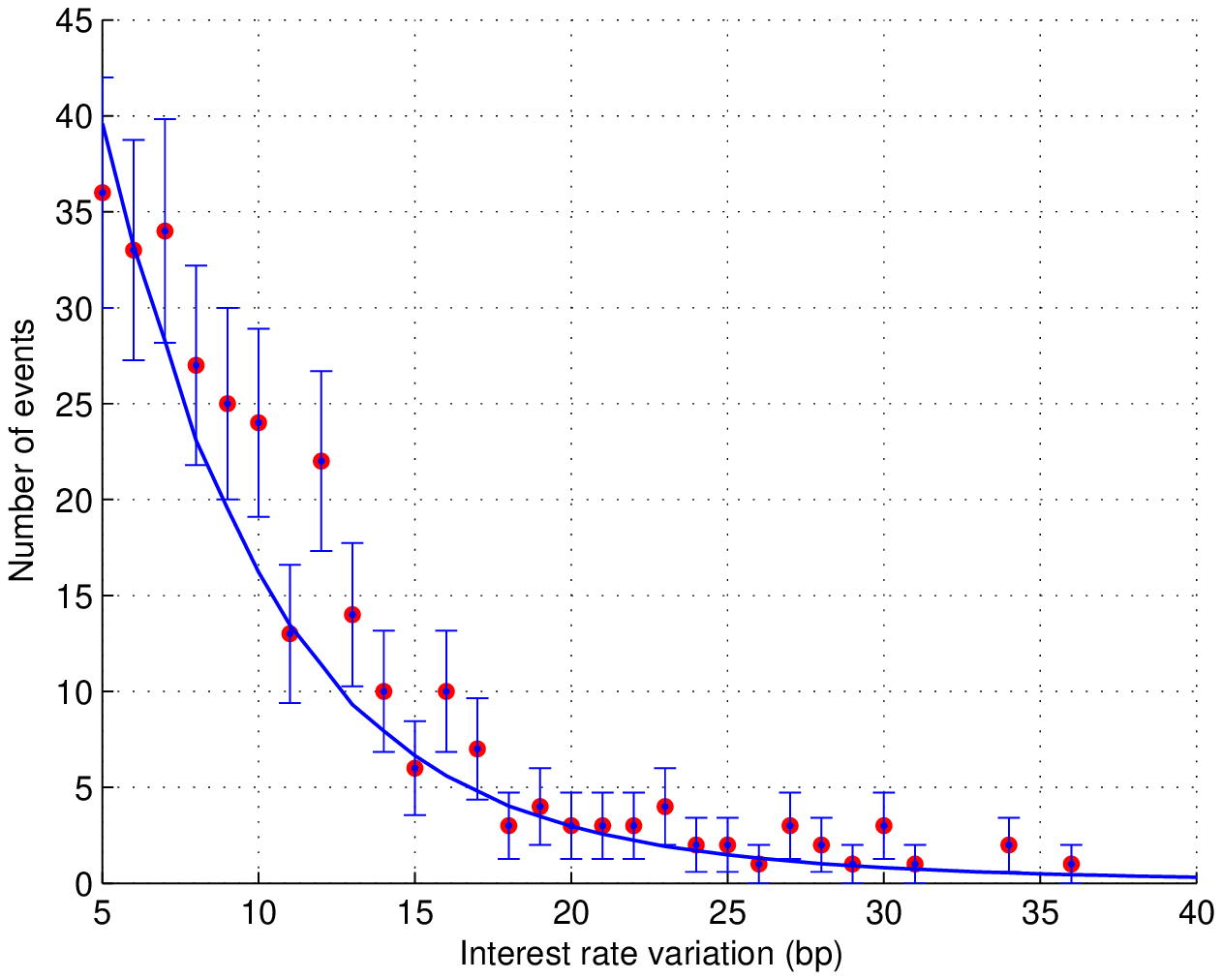}
\end{center}


\centerline{\bf{Figure 10}}
\vskip 0.3 true cm
\centerline{Maturity=$[5]$, Initial Interest Rate $\approx 5$, Lag=2}
\begin{center}
\epsfxsize=10cm\epsffile{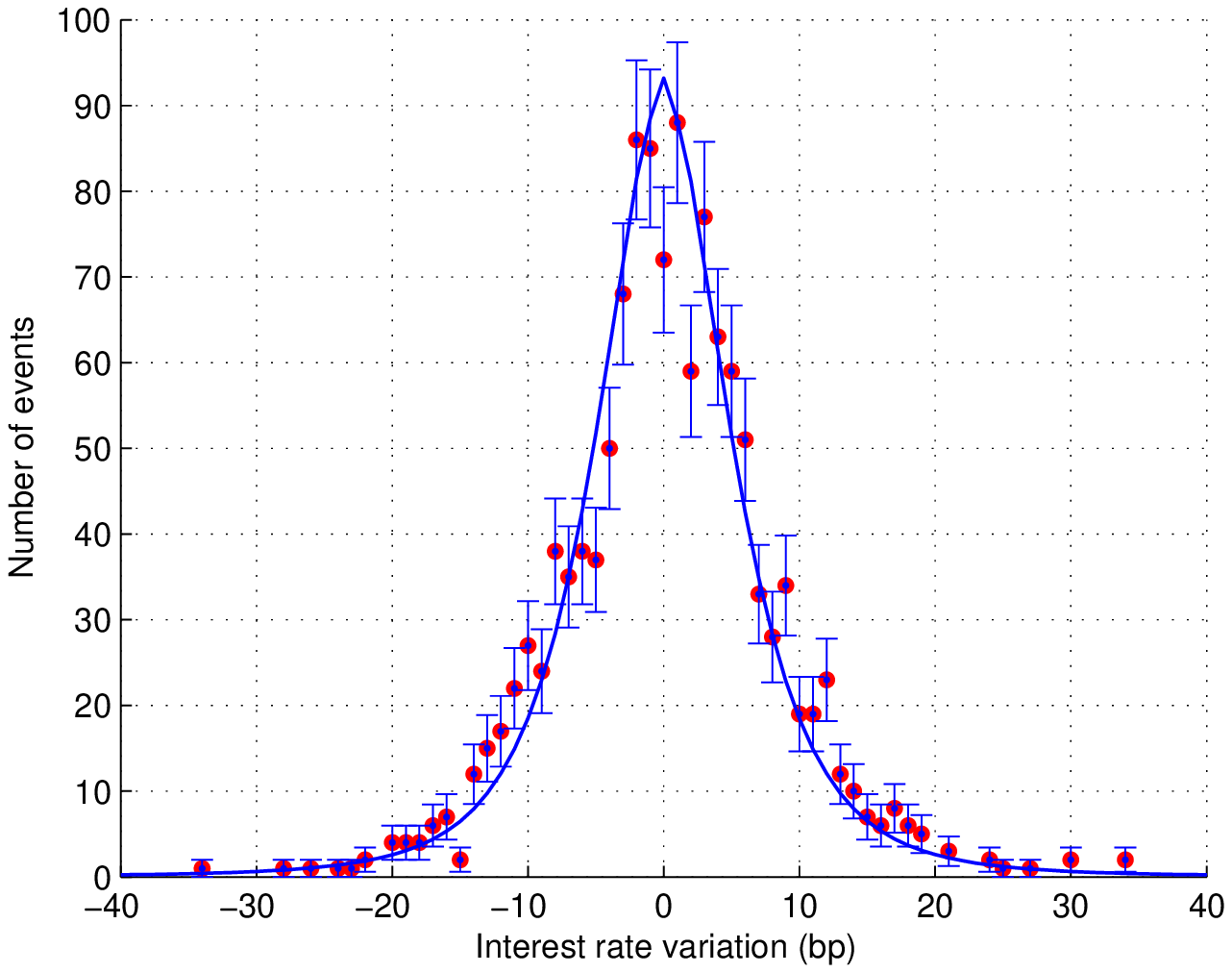}
\end{center}

\newpage

\centerline{\bf{Figure 11}}
\vskip 0.3 true cm
\centerline{Maturity=$[1]$, Initial Interest Rate $\approx 7$, Lag=2, Tail}
\begin{center}
\epsfxsize=10cm\epsffile{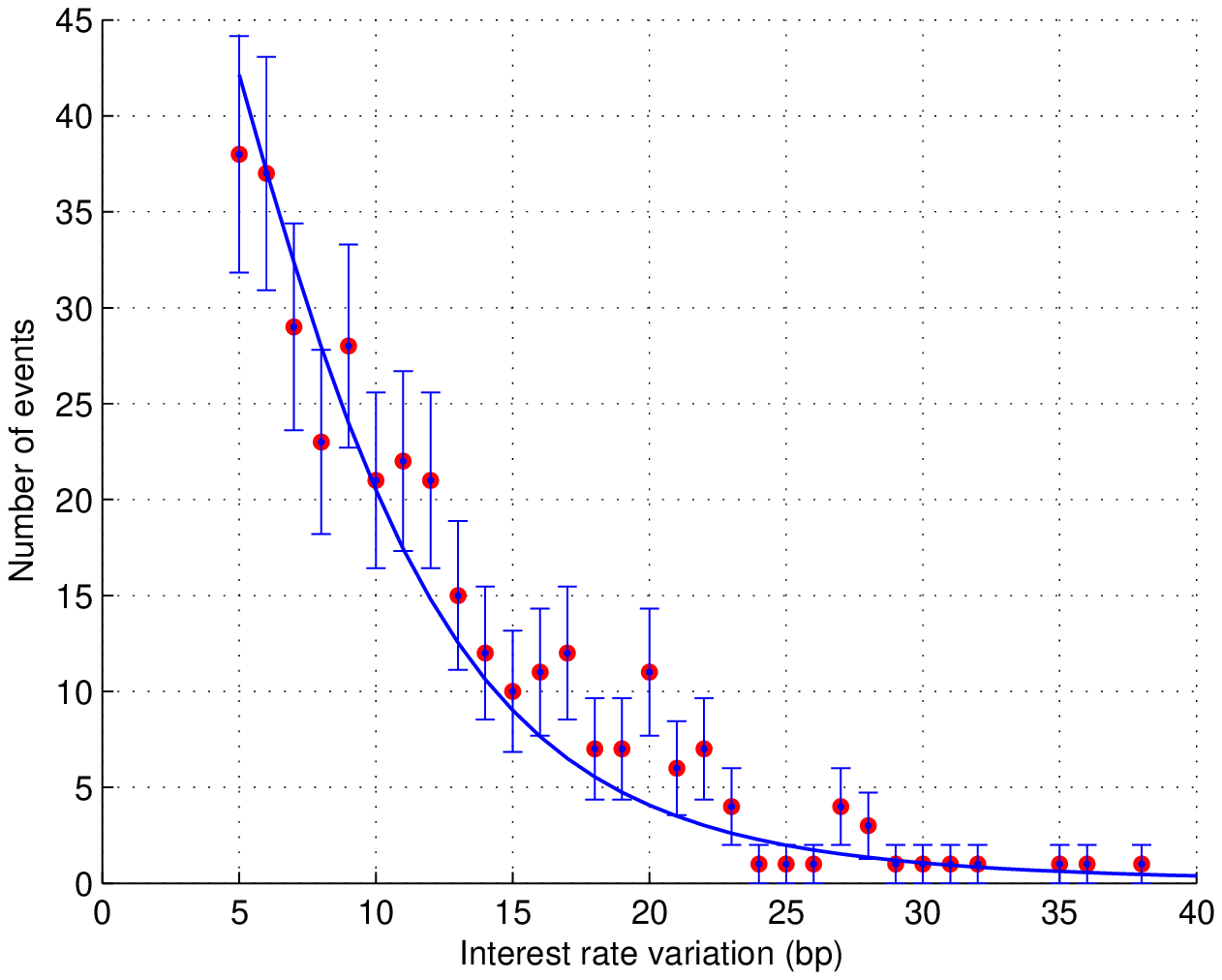}
\end{center}

\end{document}